\documentclass[aps,twocolumn,prf,nofootinbib]{revtex4}
\usepackage[utf8]{inputenc}
\usepackage{graphicx}
\usepackage{latexsym}
\usepackage{amsmath}
\usepackage{amssymb}
\usepackage{psfrag}
\usepackage{pifont}
\usepackage{color}
\usepackage{subeqnarray}
\usepackage{lipsum}
\usepackage{bm}
\newcommand{\beq}{\begin{equation}}
\newcommand{\eeq}{\end{equation}}
\newcommand{\ba}{\begin{array}}
\newcommand{\ea}{\end{array}}
\newcommand{\bee}{\begin{eqnarray}}
\newcommand{\ec}{\end{center}}

\newcommand{\m}{\mbox{\boldmath ${\mu}$}}

\newcommand{\eee}{\end{eqnarray}}
\newcommand{\bc}{\begin{center}}

\makeatletter
\begin{document}

\title{Flexible fibers in shear flow approach attracting periodic solutions} 
\author{Agnieszka M. S\l owicka$^1$}
\author{Howard A. Stone$^2$}
\author{Maria L. Ekiel-Je\.zewska$^1$\footnote{%Corresponding author. 
Email: 
mekiel@ippt.pan.pl}}
\affiliation{[1] Institute of Fundamental Technological Research, Polish Academy of Sciences, Pawi\'nskiego 5b, 02-106 Warsaw, Poland\\ 
[2] Department of Mechanical and Aerospace Engineering, Princeton University, Princeton NJ 08544}

\begin{abstract}
The three-dimensional dynamics of a single non-Brownian flexible fiber in shear flow is evaluated numerically, in the absence of inertia. 
A wide range of 
ratios $A$ of bending to hydrodynamic forces 
and hundreds of initial configurations are considered. We demonstrate that flexible fibers in shear flow exhibit much more complicated evolution patterns than in the case of extensional flow, where transitions to higher-order modes of characteristic shapes are observed when $A$ exceeds consecutive threshold values. 
In shear flow, we identify the existence of an attracting steady configuration and 
different 
attracting 
periodic 
motions that are approached by long-lasting  
rolling, tumbling and meandering dynamical modes, respectively. We demonstrate that the final stages of the first and second modes are effective Jeffery orbits, with the constant parameter $C$ replaced by an exponential function that either decays or increases in time, respectively,  corresponding to a systematic drift of the trajectories. In the limit of $C\rightarrow0$, the fiber aligns with the vorticity direction and in the limit of $C\rightarrow \infty$, the fiber periodically tumbles within the shear plane. For moderate values of $A$, a three-dimensional meandering periodic motion exists, which corresponds to intermediate values of~$C$. 
Transient, close to periodic oscillations 
are also detected in the first stages of the modes. 
\vspace{-0.3cm} 
\end{abstract}

\date{\today}
\maketitle
%r
%\vspace{-0.5cm} 
\section{Introduction}
In nature and modern technologies, 
there are many 
systems containing elongated, flexible, micrometer- and nanometer-scale objects deforming and moving in a fluid flow \cite{duRoure}. Examples are micro-swimmers such as bacteria and their flagella, actin \cite{Kantsler,Harasim2013,Liu}, large proteins or DNA molecules \cite{Chu}, and micro- or nano-fibers \cite{Lindner_2010,Nunes_2012,lindner_2015,perazzo,pawlowska}. 
Flexibility leads to complex dynamics 
with buckling \cite{Hinch1976}, coil-stretch transitions \cite{Kantsler,deGennes,BeckerShelley,YoungShelley}, migration across the streamlines of a flow \cite{graham2004,graham2005,graham2006,Slowicka2012,Slowicka_2013,farutin,Misbah2019}, knotting \cite{yeomans,stone_2015,doyle,gruziel} and a variety of deformed shapes \cite{Arlov,Skjetne,Joung,tornberg,fauci1,fauci,wang2013three,zhang2019dynamics,rost2019effective}. Similar complexity of dynamics has been observed for flexible filaments in electrokinetic fields \cite{Doyle2016} or sedimenting under gravity \cite{Lagomarsino,Netz,Llopis,Li,Gompper,BukowickiGruca,Bukowicki,Duprat2018,Bukowicki2,Gruziel2}. 

A basic question in all of these configurations is how the dynamics and shapes of deformable elongated objects in flow depend on their flexibility. 
This problem has been  investigated extensively at macro- and nano-scales for fibers in extensional, cellular and corner  flows \cite{Kantsler,YoungShelley,lindner_2015,Autrusson}. For example, it has been shown that the typical pattern  of a fiber's evolution in extensional flow is 
related to consecutive threshold values of the characteristic ratio %$A$ 
of bending to hydrodynamic forces exerted by the fluid flow. 
When these values are exceeded, higher-order modes of the fiber shape are activated, with shorter characteristic length scales of elastic deformation  \cite{Kantsler,Harasim2013,Chu,BeckerShelley,YoungShelley,Liu}. In extensional flow, there exists  a family of characteristic, well-defined  shapes.

\begin{figure}[b] \vspace{-0.1cm} 
 \includegraphics[width=10.2cm]{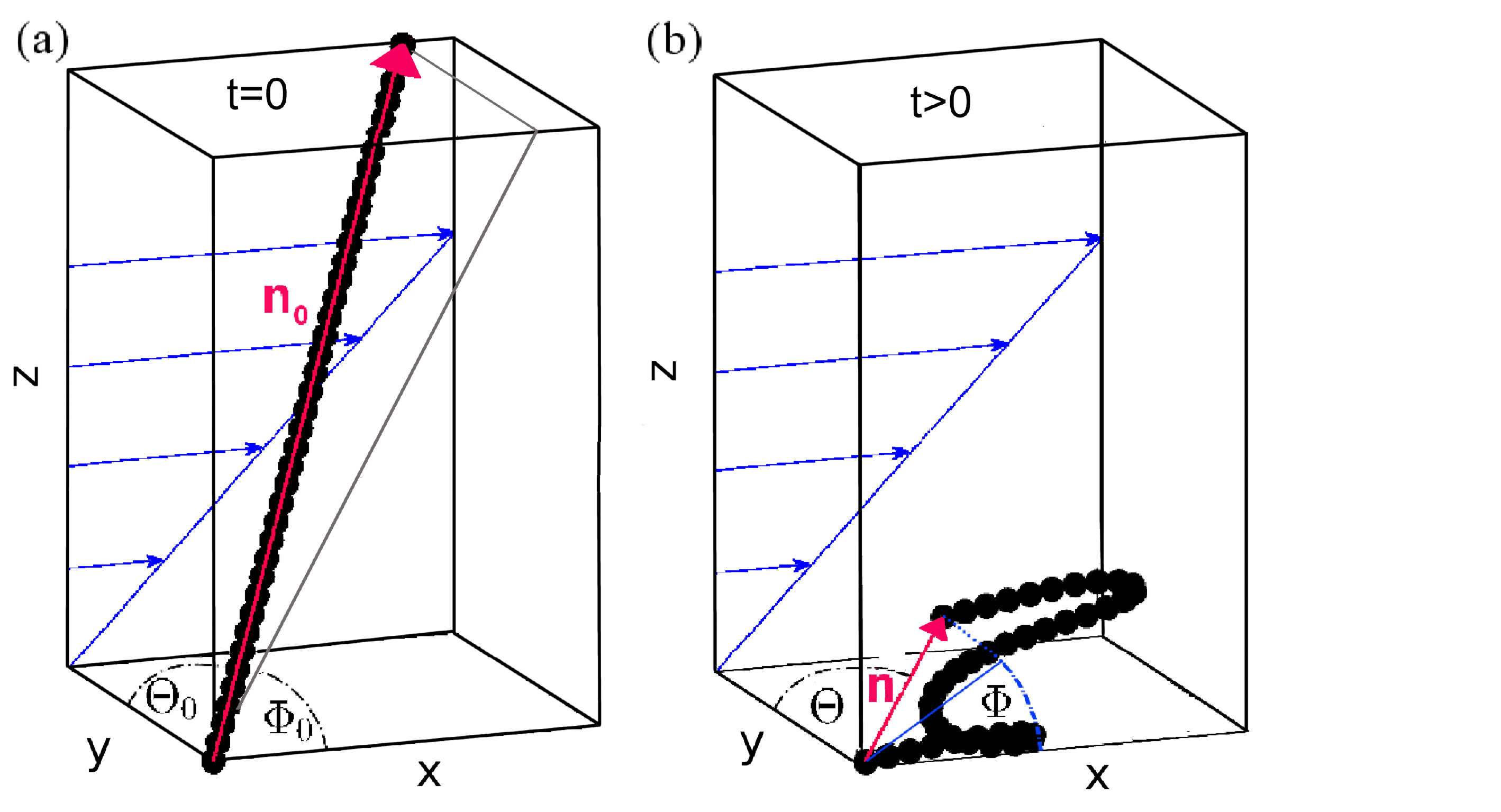} \vspace{-0.65cm}  
 \caption{Evolution of a flexible fiber in shear flow: notation.}\label{not} \vspace{-0cm}
\end{figure}

However, in general, the fiber deformation that occurs may depend on the type of flow \cite{Chu,deGennes,ICTAM}. 
Shear flows  typically occur owing to the presence of container walls, and so are of practical and fundamental interest. Therefore, as illustrated in Fig.~\ref{not}, in this paper we study three-dimensional dynamics of a single flexible fiber immersed in
steady shear flow with velocity 
\beq
{\bf v}_0\!=\!\dot{\gamma} z {\bf e}_x,\label{shear}
\eeq
where ${\bf e}_x$ is the unit vector along $x$ and $\dot{\gamma}$ is the shear rate. The Reynolds number of the system is assumed to be much smaller than unity and the fluid flow satisfies the quasi-steady Stokes equations. 

In this article we provide a new perspective on the three-dimensional evolution of flexible fiber shapes in shear flow. 
%While
We show that after a relaxation phase, a flexible fiber is attracted to one of several stationary, periodic or close to periodic solutions, with different typical sequences of shapes and orientations. Features of these characteristic solutions, their presence or absence, stability or instability, and basins of attraction depend on the ratio $A$ of local bending $E \pi d^2/64$ 
to hydrodynamic $\pi \eta \dot{\gamma} d^2$ forces, where 
$E$ is the Young's modulus, $d$ is the fiber diameter and $\eta$ is the fluid's dynamic viscosity. Therefore, by changing 
\beq
A=E/(64 \eta \dot{\gamma}),\label{AE}
\eeq
different structures and patterns of motion of dilute suspensions of flexible fibers can be obtained. For a fiber of length $L$ and diameter$d$ of the cross-section, the local bending-to-hydrodynamic ratio $A$ is related to the elasto-viscous number 
\beq
\bar{\mu}=8 (L/d)^4/A, 
\eeq
which depends on the fiber's aspect ratio $L/d$ \cite{Liu}.

A rigid straight fiber in a low-Reynolds-number shear flow follows a periodic Jeffery orbit \cite{Jeffery} (th
In this article we demonstrate that for a wide range of parameters and times, 
%provide a new perspective on the three-dimensional evolution of fiber shapes in shear flow. e specific orbit depends on the initial orientation through a constant parameter $C_0$). 
%While 
the motion and deformation of a flexible fiber can be interpreted as an effective Jeffery orbit that systematically drifts in time, owing to the exponential growth or decrease of the time-dependent parameter $C$.  For moderate values of $A\approx $ 9-12, there exists a range of values of $C$ that correspond to periodic or close to periodic meandering motions. 

A significant finding described below is that for a certain range of small values of $A \approx$ 4-5, fibers typically tend to align with the vorticity of the undisturbed flow. This result indicates a possibility to observe experimentally an ordered, dilute suspension of flexible fibers where all fibers are straight and aligned with the vorticity direction. We suggest that this ordered phase could be produced by adjusting the shear rate to reach the appropriate range of $A$. We are not aware of any such experimental observations. 

As we document in this article, in shear flow (in contrast to extensional flow), the value of the bending-to-hydrodynamic ratio $A$ does not uniquely determine the type of the fiber shape. This conclusion is based on two main features of the dynamics. First, for the same value of $A$, depending on the initial configuration or orientation, an elastic fiber evolves to a different characteristic sequence of shapes. Moreover, long-lasting, chaotic transients are typical: we document close to periodic motions that later spontaneously change into periodic or effective Jeffery motions with  different shape sequences.

\section{Theoretical model of a fiber in flow}
\subsection{Elastic fiber}
The fiber is modeled  \cite{Stark} as a chain of $N$=40 spherical beads of diameter $d$. The time-dependent position of the 
center of a bead $i$ is denoted as  %located at a position
$\bm{r}_i$. The centers  of the consecutive beads are
connected by springs of the equilibrium length $\ell_0\!=\!1.02 d$ and the potential energy  
\beq
E_s\!=\!\frac{\hat{k}}{2}\sum_i (\ell_i\!-\!\ell_0)^2.
\eeq
Here $\hat{k}$ is the spring constant and $\ell_i=|\bm{r}_i-\bm{r}_{i-1}|$ is the distance between centers of beads $i$ and $i\!-\!1$. 
In this paper we assume that the dimensionless elastic resistance ratio ${k}\!=\!{\hat{k}}/{(\pi \eta d\dot{\gamma})}$ is large, $k\!=\!1000$, which leads to an
almost constant fiber length.

At equilibrium, the fiber is straight; its deformation costs energy 
\beq
E_b\!=\!\frac{\hat{A}}{2\ell_0}\sum_i (\hat{\bm t}_{i+1}\!-\!\hat{\bm t}_i)^2, 
\eeq
dependent on the bending resistance 
\bee
\hat{A}\!=\!E \pi d^4/64. 
\eee
Here  $\hat{\bm t}_i$ is the unit vector parallel to the relative positions $\bm{r}_i-\bm{r}_{i-1}$ of the centers of beads $i$ and $i\!\!-\!\!1$. 

The total external force $\bm{F}_i$ acting on bead $i$ is elastic,
\bee
\bm{F}_i = - \frac{\partial}{\partial \bm{r}_i}(E_s+E_b).
\eee

We assume that the dimensionless 
bending-to-hydrodynamic force ratio (relative bending stiffness)
$A\!=\!{\hat{A}}/{(\pi \eta d^4 \dot{\gamma})}$, given by Eq. \eqref{AE},
varies in the range of moderate values $4\! \le \! A \!\le \!40$, where most of the fibers subsequently deform and straighten while tumbling \cite{Slowicka_chaos_2015}. %In the simulations t
The length and time units in the simulations are, respectively, $d$ and~$1/\dot{\gamma}$. 

\subsection{Initial fiber configurations}
To study the fiber evolution, we analyze the time dependence of the end-to-end vector ${\bf n}(t)\!=\!(\delta x, \, \delta y,\,\delta z)$, shown in Fig.~\ref{not}(b). We 
%denote the vector length as $\Delta L(t)=|{\bf n}(t)|$, and 
 parameterize it %orientation 
 by the standard spherical coordinates: the vector length  $\Delta L(t)=|{\bf n}(t)|$, the 
 angle $\Theta(t)$ between ${\bf n}(t)$ and the vorticity direction $y$,  and the angle $\Phi(t)$  between the projection of ${\bf n}(t)$ 
on the $xz$ plane and the $x$ axis. Initially, ${\bf n}(0)\!=\!{\bf n}_0$, $\Theta(0)\!=\!\Theta_0$ and $\Phi(0)\!=\!\Phi_0$.

To investigate the characteristic features of the flexible fiber dynamics in shear flow, we consider the following  family of initial conditions. We assume that at $t=0$ the fiber is straight and that all of the springs between the beads are at their equilibrium lengths \cite{ICTAM}. % \cite{Arlov,Forgacs,Skjetne,Joung,Wang_Yu,ICTAM}. 
The initial orientation of the fiber is given by the orientation of the end-to-end vector ${\bf n}_0$ that links the centers of the first and the last beads.  The length of this vector is  $|{\bf n}_0 | %= \Delta L(0)
\!=\!L_0\equiv (N-1)\ell_0$. The direction of 
${\bf n}_0$ is parameterized by the spherical angles $\Theta_0$ and $\Phi_0$, as indicated in Fig.~\ref{not}(a). 
We consider the whole range of the initial orientations.

In the following, we will systematically investigate how the fiber's dynamics and shape evolution depend on the relative bending stiffness $A$ and the initial orientation $(\Theta_0,\Phi_0)$. In this way we will study the three-dimensional dynamics of a flexible fiber in shear flow, while most of the previous studies %of flexible fibers motion 
have focused on the two-dimensional dynamics in the shear plane. % \cite{........................}. 

\subsection{%Theoretical description of the 
Fiber dynamics in flow}%the moving fluid}% in Hydrodynamic interactions}
The dynamics of an elastic fiber is determined by the external (i.e.,~elastic) forces exerted on each fiber bead, and the hydrodynamic interactions between them caused by the presence of the shear flow. We assume that the dynamic viscosity $\eta$ of the fluid that surrounds the fiber is large enough for the Reynolds number to be much less than unity. In this limit, the fluid 
velocity  $\bm{v}$ and pressure \textit{p} satisfy the 
Stokes equations \cite{batchelor1967,kim2013microhydrodynamics}, \bee %\begin{equation}
\eta \nabla^2 \bm{v} - \nabla p   =  \bm{0},  \eee
The 
%where $\eta$ is the fluid dynamic viscosity, %, and $\bm{f}$ is an external body force exerted on the fluid element of unit volume. 
%The Stokes equations correspond to low-Reynolds-numbers  \cite{batchelor1967,kim2013microhydrodynamics}. % $\frac{D}{Dt}$ is the time derivative following the fluid element, $\rho$ is the fluid density, 
%Appropriate
no-slip boundary conditions 
at the beads surfaces are assumed. The fluid is unbounded, with the %shear 
ambient flow velocity ${\bf v}_0$ given by Eq.~\eqref{shear}; in the presence of the fiber, the fluid velocity $\bm{v}$  tends to ${\bf v}_0$ when the distance from the fiber goes to infinity. 

%pure shear at the surfaces confining the fluid (or at infinity) of solving these equations are necessary to investigate dynamics of micro-particles and to determine structure and effective transport properties of dispersive systems. 

%The motions of the beads are determined by performing the multipole expansion, corrected for lubrication effects \cite{cichocki}, and implemented in the {\sc Hydromultipole} numerical code.  The method is 
%{\bf Theoretical description of microparticle dynamics in fluids} of viscosity $\eta$ is
%\subsubsection{Particle motion: multipole expansion corrected for lubrication}
To solve the Stokes equations in the presence of the fiber, we  use %simple analytic representations or (alternatively) 
the advanced theoretical algorithm \cite{cichocki1994friction,ekiel2009precise}, based on the  multipole expansion corrected for lubrication \cite{cichocki}, implemented in numerical codes {\sc Hydromultipole} 
(\cite{cichocki}). The method, similar to \cite{Brady,Ladd}, is based on the boundary integral representation of the fluid velocity and the boundary integral equations for the surface density of the forces  induced at the particle surfaces. These equations are projected on a complete set of elementary solutions of the Stokes equations (spherical multipole functions). The resulting set of linear algebraic equations is truncated at a controlled multipole order.%~$L$.
 The convergence of the multipole expansion is speed up by applying the lubrication correction \cite{cichocki,Bossis,Sangani}.
%The {\it generalized (grand) mobility matrix} consists of the consequtive multipole elements $\m^{pq}$, with $p,q=t,r,d,...$, and depends on the configuration of all the particles.
%
%
%The main advantage of the Hydromultipole codes is high efficiency and controlled accuracy, by a suitable choice of $L$. %, which is controlled by the choice of the multipole order of the truncation. This theoretical algorithm was implemented numerically into the {\sc Hydromultipole} codes, which 
%They have been extensively used  for more than twenty years  
%\cite{felderhof1976force,cox1967force,mazur1974force,abade2010short,pasol2011motion, farutin2016dynamics}.

In general, dynamics of $N$ spherical beads moving %under %,, 
in an ambient Stokes flow have the form 
\bee
\left( \ba{cc}
\bm{U} \\
\bm{\Omega}
\ea
\right) 
 &=& \left( 
\ba{cc}
{\m}^{tt}&{\m}^{tr}\\
{\m}^{rt}&{\m}^{rr}\\
\ea
\right) \cdot
\left( \ba{cc}
{\boldsymbol{F}+\boldsymbol{F}_0 }\\
{\boldsymbol{{\cal T}}+\bm{{\cal T}}_0}
\ea
\right), \label{flows} 
\eee
where 
$\bm{U}\!\!=\!\!(\bm{U}_1,...,\bm{U}_N)$ and $\bf \Omega\!\!=\!\!(\bf \Omega_1,...,\bf \Omega_N)$ are 
 translational and rotational bead velocities, 
$\bm{F}\!\!=\!\!(\bm{F}_1,...,\bm{F}_N)$ and $\bm{{\cal T}}\!\!=\!\!(\bm{{\cal T}}_1,...,\bm{{\cal T}}_N)$ are 
 external forces and torques (couples) exerted on the particles; 
$\bm{F}_0\!\!=\!\!(\bm{F}_{0,1}...,\bm{F}_{0,N})$ and $\bm{{\cal T}}_0\!\!=\!\!(\bm{{\cal T}}_{0,1},...,\bm{{\cal T}}_{0,N})$
are forces and torques exerted by the ambient flow on the motionless beads. Here, $\bm{\mu}_{ij}^{tt},\;\bm{\mu}_{ij}^{tr},\;\bm{\mu}_{ij}^{rt},\;\bm{\mu}_{ij}^{rr}$ are the  3$\times$3 translational-translational, translational-rotational, rotational-translational and rotational-rotational mobility matrices, respectively. They depend on 
  positions $\bm{r}_i$ of the centers of all the beads $i=1,...,N$. 
  
  In this paper, Eq.~\eqref{flows} simplifies, because there are no external torques, $\bm{{\cal T}}\!\!=\!\!{\bf 0}$, and we are not interested in the bead rotations $\bm{\Omega}$. %To determine the motion of the fiber beads, 
The  {\sc Hydromultipole} codes %corrected for lubrication
%\cite{cichocki1994friction,ekiel2009precise}, described above, and %applying 
\cite{cichocki} are used to evaluate with high precision $\bm{\mu}_{ij}^{tt},\;\bm{\mu}_{ij}^{tr},\;\bm{\mu}_{ij}^{rt},\;\boldsymbol{F}_0,\;\boldsymbol{{\cal T}}_0$ for given positions $\bm{r}_i$ of the bead centers, and  Eqs.~\eqref{flows} are solved numerically with the adaptive fourth-order Runge-Kutta method. More information about the numerical method and its accuracy is given in Appendix \ref{C}. \\ \\
%the translational-translational and translational-rotational mobility matrices  and ............. 
%In general self-terms also depend on particle positions. 

\section{Attracting modes of the dynamics}
One of our most significant findings is that, depending on the initial orientation and bending stiffness, the 
fiber is attracted to one of three distinct stationary or periodic solutions. The characteristic properties of these evolution patterns are illustrated in Fig.~\ref{panel1} and Movie~1, for $A\!\!=\!\!10$ and three different initial orientations of the end-to-end vector ${\bf n}_0$, $(\Theta_{0},\Phi_{0})\!\!=\!\!(5^{\circ},10^{\circ})$, $(45^{\circ},30^{\circ})$, %and 
$(10^{\circ},10^{\circ})$. The colors blue, red and green indicate time-dependent modes attracted to different stationary or periodic solutions. In the following these attractors  will be called ``rolling'', ``tumbling'' and ``periodic meandering'', respectively. 
\onecolumngrid
\begin{center}
\vspace{-0.3cm}
\begin{figure}[h]% [h] 
\hspace{-0.29cm}
\includegraphics[width=18cm]{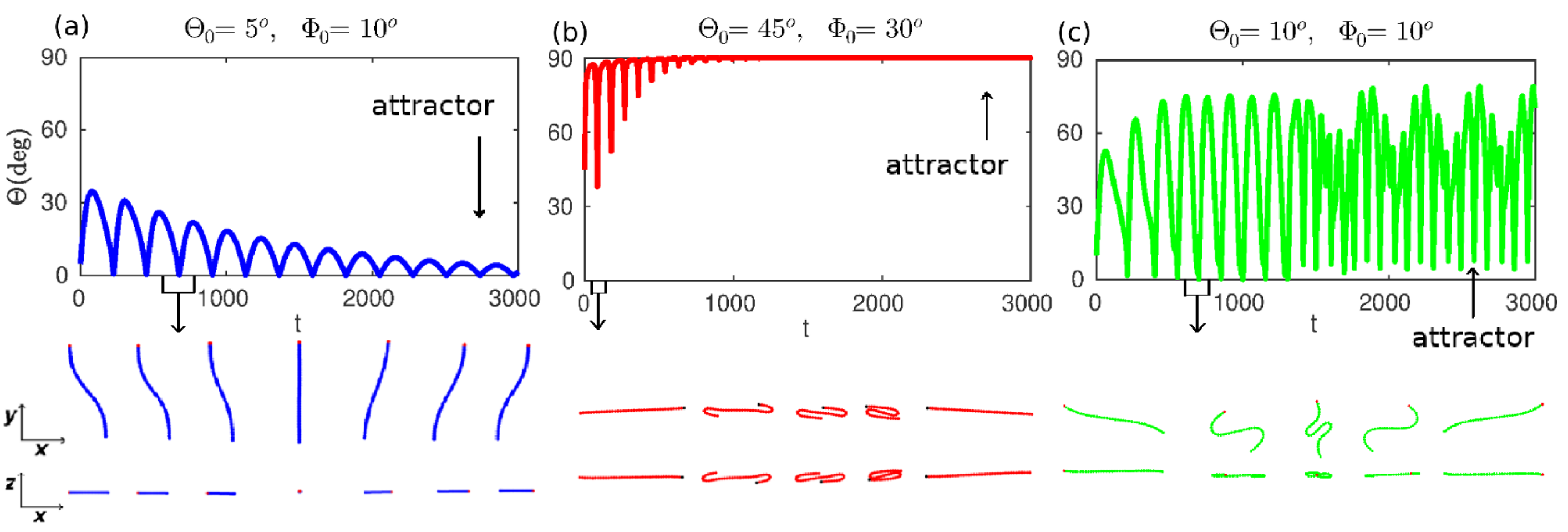}\hspace{-0.42cm}
\vspace{-0.25cm}
\caption{
Different modes of the dynamics, approaching different attractors: (a) rolling (blue online), (b) tumbling (red online), (c)~meandering~(green online). Examples shown are for  $A$=10 and different  orientations  $(\Theta_0,\Phi_0)$ of the initial end-to-end vector ${\bf n}_0$  as indicated. The orientation angle~$\Theta$ of the end-to-end vector ${\bf n}$ is plotted versus time. Shape evolution of the fibers is also shown in the $xy$ and $xz$ projections for the time window indicated. In (c) the transient squirming (close to periodic) 
sequence of shapes is shown, approximately repeating after 
$t\! \approx \!305$. Snapshots are taken at (a) $t\!\!=\!\!640, 675,710,759, 808, 862, 885$; (b) $t\!\!=\!\!50, 78, 84, 90, 118$; (c) $t\!\!=\!\!615,660,690,720,766$.
\vspace{-0cm}
}
\label{panel1}
\end{figure}%
\end{center}
%\twocolumngrid
%
%\twocolumngrid
%\newpage
%\onecolumngrid
\begin{figure}[t!]
%14.07
\bc
\vspace{-0.6cm}
\includegraphics[width=16.5cm]{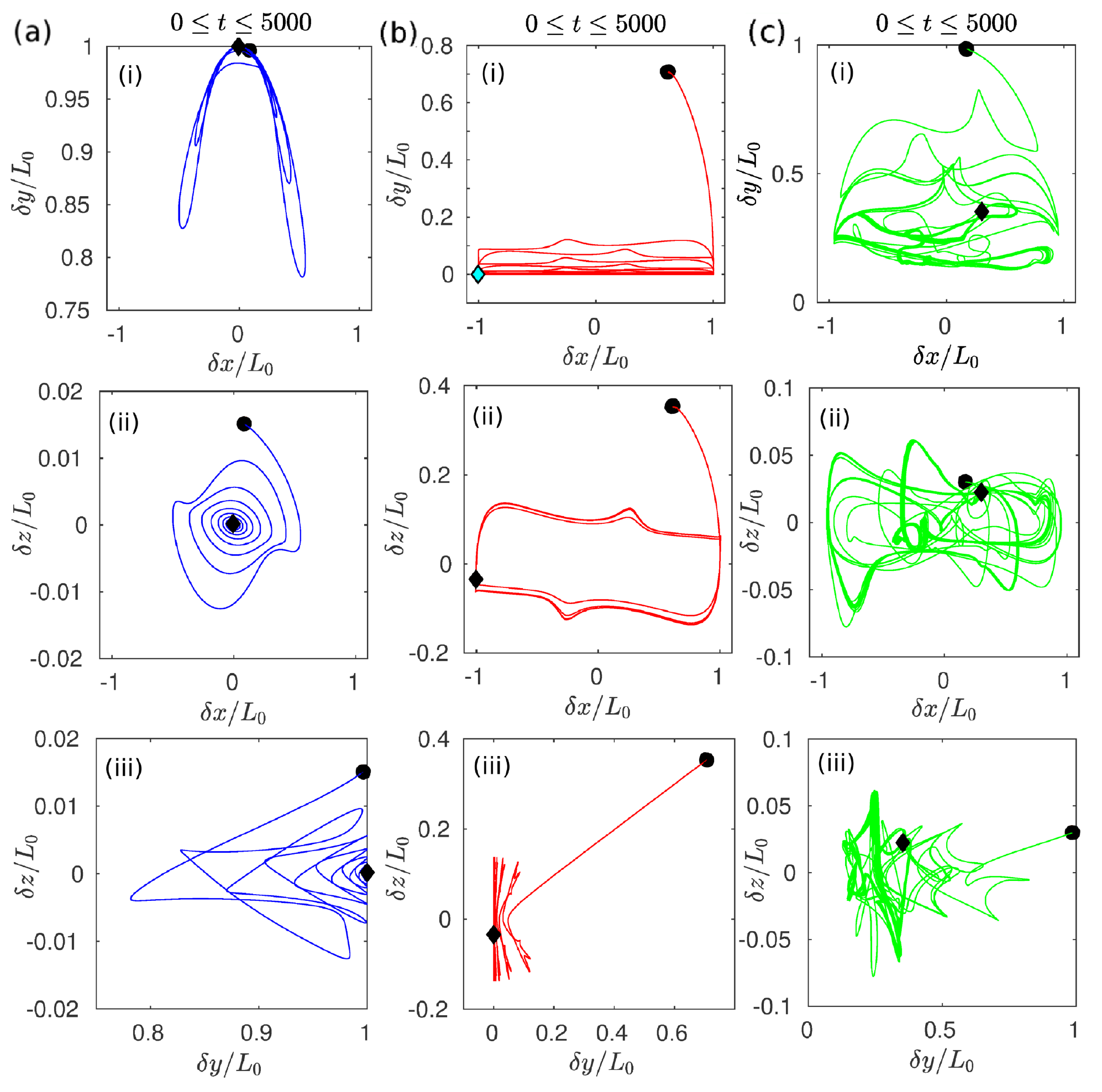}  
 \vspace{-0.5cm}
\caption{Three evolving modes. Trajectories of the fiber end-to-end vector for the three modes with $A$=10, shown in Fig.~\ref{panel1}: (a) rolling (blue online), (b) tumbling (red online) and (c) meandering (green online). A given initial condition is attracted to one of  three characteristic orbits (rolling, tumbling and meandering), shown in %Fig. \ref{nn} and 
Movie~2 and Sec. \ref{attractors}. Symbols: ${\Huge \bullet}\; %brown dot, 
t$=0; \;%cyan 
%diamond,
$\blacklozenge\;t$=5000.} \label{panel2add} \vspace{-1.05cm}
\ec
%Characteristic properties of different periodic modes for fibers starting from close inital positions.
%%%Trajectories of the fiber end-to-end vector in the final stage of the evolution, $t_{end}\!-\!800 < t < t_{end}$, with $t_{end}\!=\!15000$ for the blue and red modes, $t_{end}\!=\!100000$ for the green mode.}
\end{figure}
%\end{widetext}\\
\twocolumngrid
The rolling mode (also called spin-rotation \cite{Skjetne}, or log rolling) is shown at early times in  Fig.~\ref{panel1}(a) and Movie 1. In this mode, the fiber end-to-end vector tends at long times to the vorticity ($y$) direction, i.e. $\Theta \!\rightarrow\!0$. Moreover, all of the fiber beads rotate around $y$, and tend to align along $y$. The tumbling mode is shown at early times in Fig. \ref{panel1}(b) and Movie 1. In this mode, the fiber end-to-end vector tends at long times to the $xz$ plane ($\Theta\! \rightarrow \!90^{\circ}$).   The convergence is confirmed by much longer simulations with $0 \!\le\! t \!\le \!15000$.

The attracting character of the rolling and tumbling 
solutions have been identified previously  \cite{Arlov,Forgacs,Skjetne,Joung,Wang_Yu,ICTAM} as the generic 
feature of the flexible fiber dynamics, and the corresponding dynamical modes have been called by the same names as their attractors \cite{Skjetne,Joung,stone_2015}.

To our surprise, we discovered that there exists a third dynamical evolution, illustrated in  Fig.~\ref{panel1}(c) %-\ref{shapes}, 
and Movie~1 
%-2, and Figs. 9-11 in Supplemental Material 
-- an approach to an attracting three-dimensional ``periodic meandering''  solution, with $\Theta$ oscillating periodically for $1800 \! \lesssim \! t \! \le \!10^5$. To the best of our knowledge, this behavior has not been observed before.

 %The rolling mode (also called spin-rotation \cite{Skjetne}, or log rolling) is shown at early times in  Fig.~\ref{panel1}(a) and Movie 1. In this mode, the fiber end-to-end vector tends at long times to the vorticity ($y$) direction, i.e. $\Theta \!\rightarrow\!0$. Moreover, all of the fiber beads rotate around $y$, and tend to align along $y$. The tumbling mode is shown at early times in Fig. \ref{panel1}(b) and Movie 1. In this mode, the fiber end-to-end vector tends at long times to the $xz$ plane ($\Theta\! \rightarrow \!90^{\circ}$). 

\begin{figure*}[ht!]
\vspace{-0.2cm}
\bc
%14.07
%\includegraphics[width=13.8cm]{figures/figure11.eps}
%\includegraphics[width=23.2cm]{fig7.eps}
%\includegraphics[width=17.2cm]{Figure8.eps}
%\includegraphics[width=17.2cm]{fig13n.eps}
\includegraphics[width=17.2cm]{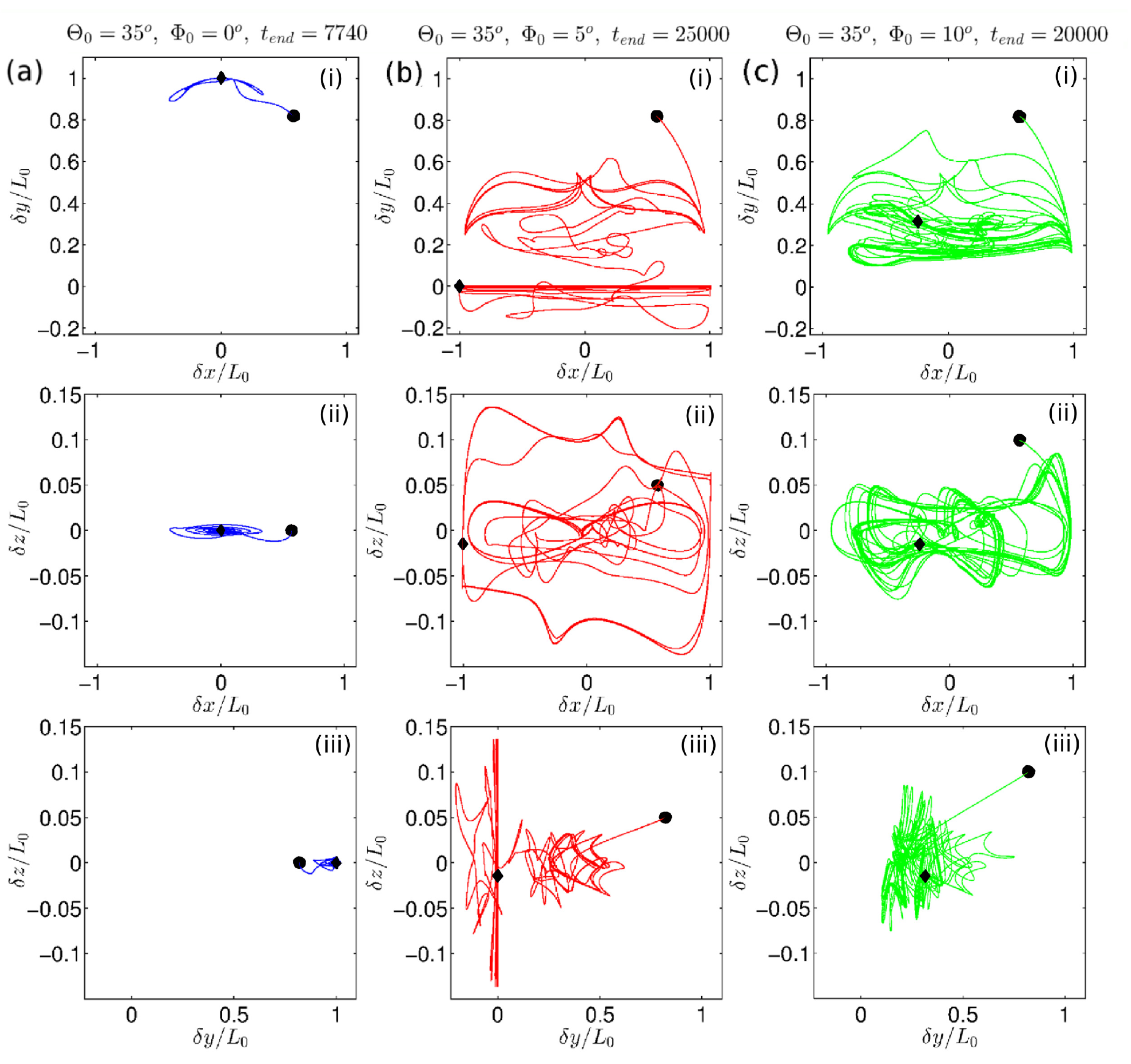}
%figures/figure11a.eps}  
\vspace{-0.44cm}
\caption{Sensitivity to a small change of the fiber initial orientation (indicated by the dot in each panel). The initial orientation in each column is the same, and there is only a small difference in initial orientations in the different columns. Each initial orientation %(as indicated) 
is attracted to a different orbit: (a) rolling (blue online), (b) tumbling (red online) and (c) meandering (green online). 
The trajectories of the fiber end-to-end vector are shown for  $A$=10 and $0\!\le t \le \!t_{end}$. %, displayed in column (d) with the same colors. 
Symbols: ${\Huge \bullet}$ %brown dot, 
$t$=0; \;%cyan 
%diamond,
$\blacklozenge\;t=t_{end}$, with $t_{end}=15000$ in (a) and (b) and  $t_{end}=10^5$ in (c). 
%Symbols: brown dot, $t$=0; cyan diamond, $t_{end}$ as indicated.
%%%%%%%%%=5000; black open square, $t_{end}$=99200. 
The tumbling period $T_t = 181$ and the meandering period $T_m=735$. \vspace{-0.55cm}
%Characteristic properties of different periodic modes for fibers starting from close inital positions.
}
%%%Trajectories of the fiber end-to-end vector in the final stage of the evolution, $t_{end}\!-\!800 < t < t_{end}$, with $t_{end}\!=\!15000$ for the blue and red modes, $t_{end}\!=\!100000$ for the green mode.}
\label{panel_t35}
\ec
\end{figure*}

The results in Fig.~\ref{panel1} are focused on the relaxation phase after which the attractors are reached. %simulations went on much longer, confirming the shown behavior. 
The relaxation time  can be very long, of the order of thousands of dimensionless units, as can be seen in Fig.~\ref{panel1}. Moreover, long-lasting, close to periodic, transient motions (different from tumbling and meandering), which we will call ``squirming'', can appear in the relaxation phase, as shown in Fig.~\ref{panel1}(c) and Movie 1.  Squirming motions are typically present for a wide range of initial orientations and values of $A$, and in all of the modes - those that tend to meandering, rolling or tumbling solutions.  
%To our surprise, we discovered that there exists a third dynamical evolution, illustrated in  Fig.~\ref{panel1}(c) %-\ref{shapes}, and Movie~1  -- an approach to an attracting three-dimensional ``periodic meandering''  solution, with $\Theta$ oscillating periodically in time. To the best of our knowledge, this behavior has not been observed before. 

The approach to three different attracting solutions is also well illustrated 
%We characterize the three modes as illustrated in Movie~3 -- 
by plotting the trajectories drawn by the tip of the end-to-end vector ${\bf n}(t)\!=\!(\delta x, \, \delta y,\,\delta z)$. This concept is explained in Movie~3. 
The results obtained for $A$=10, and the same initial orientations as in Fig.~\ref{panel1},
are shown in Fig.~\ref{panel2add}. %.... 6-8 .......in the Supplemental Material. 
%
%The approach to three different attracting solutions is well illustrated by the dynamics of the end-to-end fiber trajectory, displayed in Figs.~\ref{panel2add} (see also fig. \ref{panel_t35} in Appendix \ref{AA}. In Fig.~\ref{panel2add}, we present the results for $A$=10 and the same initial orientations as those shown in Figs. \ref{panel1}-\ref{shapes}  of the main text. 
We use different scales for different modes to illustrate the detailed characteristic features of each evolving mode, 
% Each of the modes 
which is attracted by a different solution: stationary rolling, periodic tumbling and periodic meandering, respectively. Approaching these solutions takes a long time. We highlight in Fig.~\ref{panel2add}(a)(ii) the regular, anisotropic spiraling towards the steady state. We demonstrate in Fig.~\ref{panel2add}(b) a fast convergence of ${\bf n}$ to the shear plane. 
We illustrate in Fig.~\ref{panel2add}(c) the existence of a transient, almost periodic squirming %, end-to-end
trajectory in 
the relaxation phase of the meandering mode.  %, an almost periodic %, end-to-end
%trajectory is visible, see Fig. \ref{panel2add}(c), which 
%It corresponds to the transient squirming motion, 
%(see also %visible in 
%Fig. \ref{panel1}(c)).

%It is interesting to study
The transition between the different modes is illustrated in Fig.~\ref{panel_t35}. The columns %Figs.~\ref{panel_t35}
(a)-(c) show trajectories of the fiber end-to-end vector for 
%we demonstrate evolution of the fiber with 
$A$=10 and three different initial orientations, which are very close to each other, but represent different %dynamical
modes; %By analogy to  Fig.~\ref{panel2add}, we  
we keep the same scale in each row of the figure.  %for all the initial orientations.
It is clear that the dynamics can be sensitive to a small change of the initial conditions. 

%{\bf Sensitivity .... How do different dynamical modes approach rolling, tumbling and meandering solutions?}

%
%\newpage
%%$\:$\\ \newpage

\begin{figure}[t!]
\vspace{-0.31cm}
\includegraphics[width=7.2cm]{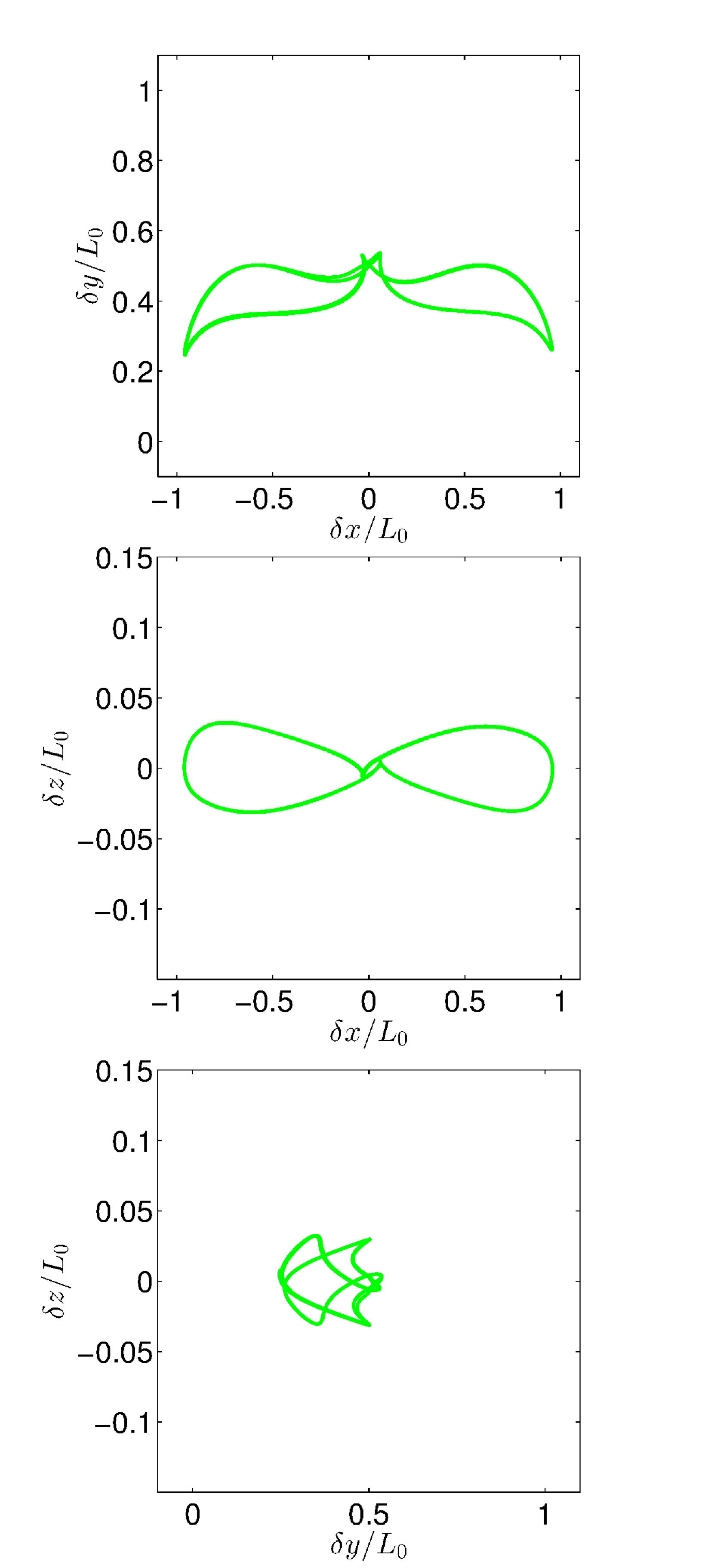} \vspace{-0.09cm}
\caption{Transient, close to periodic squirming motion for $A$=10 and the same initial conditions  as in Figs. \ref{panel1}-\ref{panel2add}(c) and Movie 2. The time range is $600 \!\!\le\!\! t \!\!\le \!\! 1224$. Trajectories of the fiber end-to-end vector are close to periodic; %consecutive 
shapes are a little different, but almost repeating after $t \approx 305$. \vspace{-0.5cm}
}\label{g2}
\end{figure}
We next analyze the trajectories in Fig.~\ref{panel_t35}(b,c) during the relaxation phase. By comparing them,  %trajectories plotted in red and green in Fig. \ref{panel_t35}(b) and (c), respectively, 
we observe that at short and moderate times there appear a range of times with the characteristic squirming trajectory, which repeats almost periodically. This squirming end-to-end trajectory is  shown  separately in Fig. \ref{g2}. 

Summarizing, the transient periodic squirming motion is visible not only in the meandering mode  (which ends up at the periodic meandering solution), but also in the tumbling mode (which ends up at the tumbling solution). % trajectory, which finally leads to the periodic tumbling solution. 
For smaller values of the relative bending stiffness $A$, squirming motion is also often visible before the rolling solution is reached. The attractors -- the rolling, tumbling and meandering solutions -- will be shown in the next section. %are shown in  Fig. \ref{panel_t35}(d). 
\section{The attractors: stationary and periodic solutions}\label{attractors}
\begin{figure} [t!]
\includegraphics[width=5.96cm]{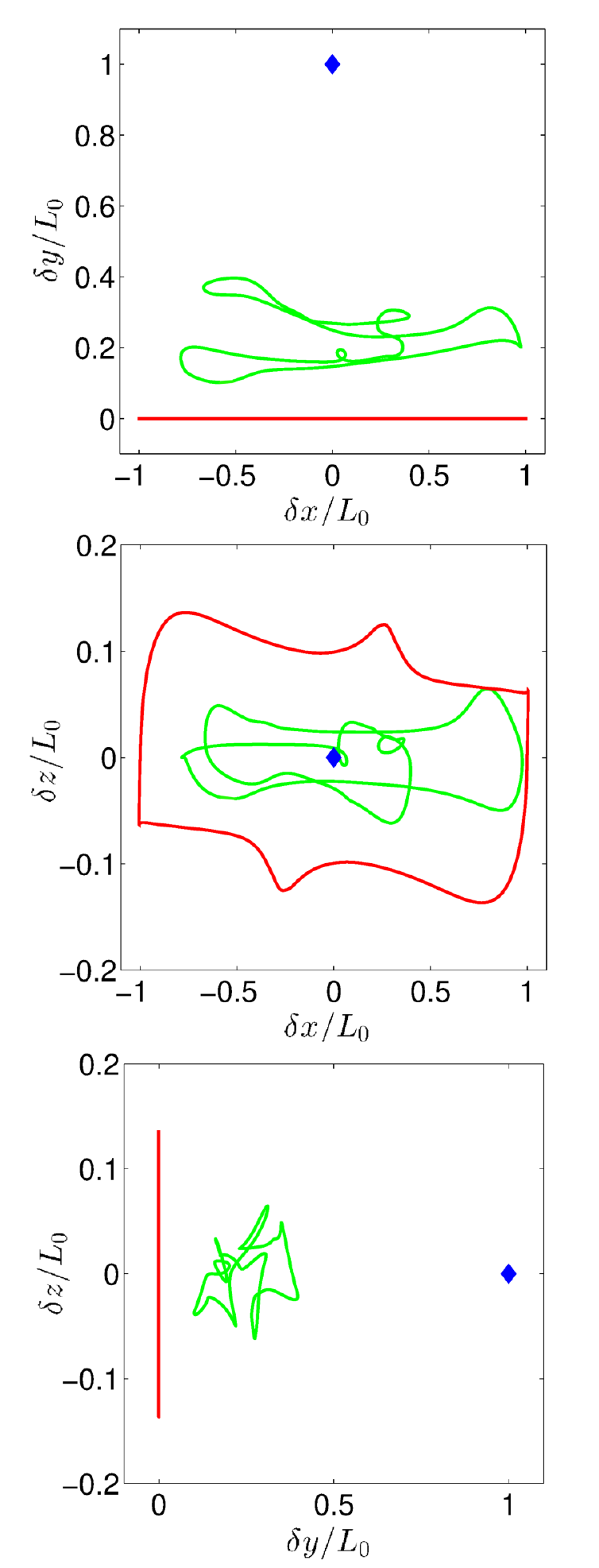} 
\vspace{-0.14cm}
\caption{Three attractors of the fiber dynamics: rolling stationary configuration %aligned with the vorticity direction
(blue diamonds), together with tumbling (red) and meandering  (green) periodic orbits. %The trajectories of the fiber end-to-end vector 
%are plotted for $14200 \le t \le 15000$ (blue and red) and $10^5-800 \le t \le 10^5$ (green). ........ (c) 
The $xy$, $xz$ and $yz$ projections of the end-to-end periodic trajectories are plotted for 
$t_{end}-800 < t < t_{end}$, with $t_{end}\!=\!10^5$ 
for meandering, and $t_{end}$=15000 
for tumbling trajectories. %The rolling stationary configuration is marked by blue dots.
}\label{nn}\vspace{-0.2cm}
\end{figure}

\begin{figure*}[ht!]
\vspace{-0.3cm}
\hspace{-0.35cm}
\includegraphics[width=18.2cm]{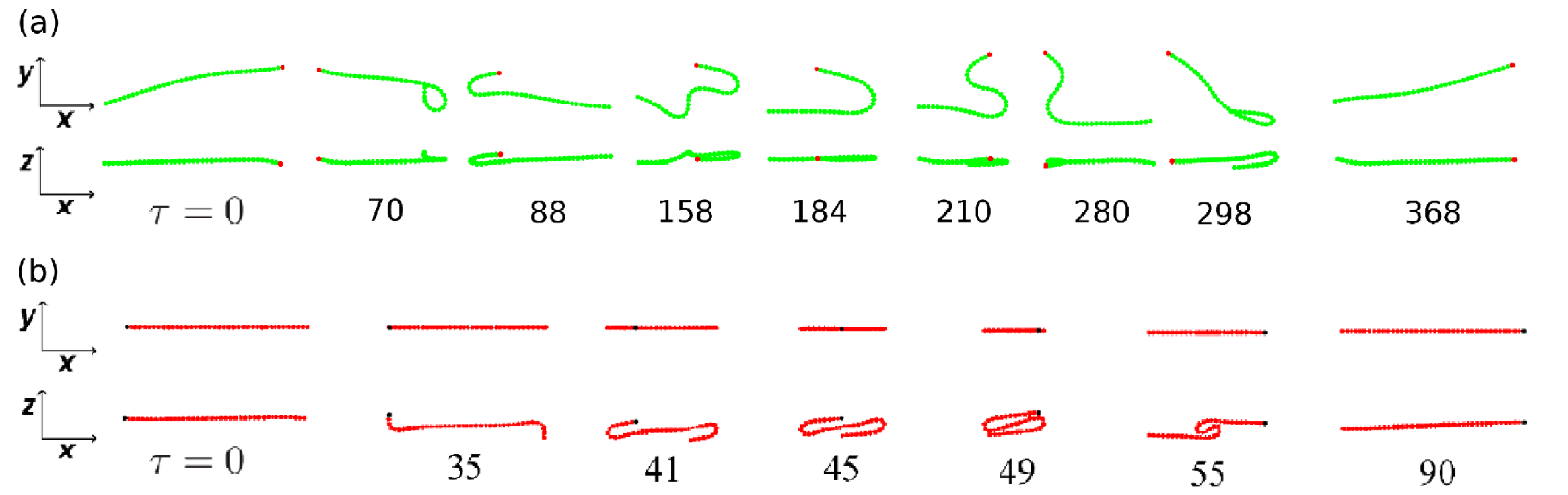} \vspace{-0.7cm}
\caption{Periodic shape sequences for the meandering (green online) and tumbling (red online) attracting periodic solutions with $A$=10, reached from  the same initial orientations as in Figs.~\ref{panel1} and~\ref{panel2add}. Snapshots are taken at  $t$=$t_0$+$\tau$, with (a) $t_0$=$99556$, (b) $t_0$= 4653, and $\tau$ as indicated. 
}
\label{shapes} \vspace{-0.4cm}
\end{figure*}

%\vspace{-0.1cm}
The three modes presented in Figs.~\ref{panel1}-\ref{panel_t35} tend to three 
different attracting solutions that are compared with each other in Figs.~\ref{nn}-\ref{shapes}. 
As illustrated in Fig.~\ref{nn}, 
the ordering of the end-to-end vector is different for each of these attractors, with the corresponding mean values $\langle \delta y/L_0\rangle$  well-separated from each other. 

%\onecolumngrid
%\begin{center}

%\end{center}
%\twocolumngrid

In the rolling solution, the fiber is straight %, with $\Delta L/L_0 \!\rightarrow \!1$, 
and it spins along the vorticity direction $y$.
In the tumbling periodic solution (with the period $T_t\!=\!181$ for $A$=10), the fiber end-to-end vector is perpendicular to $y$. It tumbles in the shear plane around the vorticity direction while the fiber straightens along the flow and then becomes coiled, with large-amplitude oscillations of length of its end-to-end vector. %$\Delta L/L_0$.

The periodic meandering orbit (with the period $T_m\!=\!735$ for $A$=10) has been observed %does not destabilize 
even in very long simulations with more than 130 periods. Details about the periodic meandering motion are given in Appendix~\ref{A}. The evolution of shapes in the meandering and tumbling  periodic solutions are %presented 
compared with each other in Fig.~\ref{shapes}(a,b) and Movie 2. In both periodic orbits, the $xz$ projections of the fibers have small amplitudes $\delta z$ along the flow gradient direction, but the corresponding shapes are essentially different from each other. Moreover, in the tumbling motion, all of the fiber beads stay in the shear plane (i.e., $\delta y$=0), while the meandering motion is three-dimensional, with large values of the end-to-end projection $\delta y$ along the vorticity direction.~Therefore, the meandering and tumbling periodic motions differ 
significantly from each other. %the Jeffery orbits \cite{Jeffery}, and 

%indicate that %, as  
%
%
%shown in Fig.~\ref{shapes} (see also Movie 2). % and Fig.~10 of the Supplemental Material). 

%\vspace{14.6cm}

\vspace{-0.1cm}
\section{Dependence of the dynamics on the bending stiffness ratio $A$}
\begin{figure}[h] \vspace{-0.2cm}
\includegraphics[width=8.9cm]{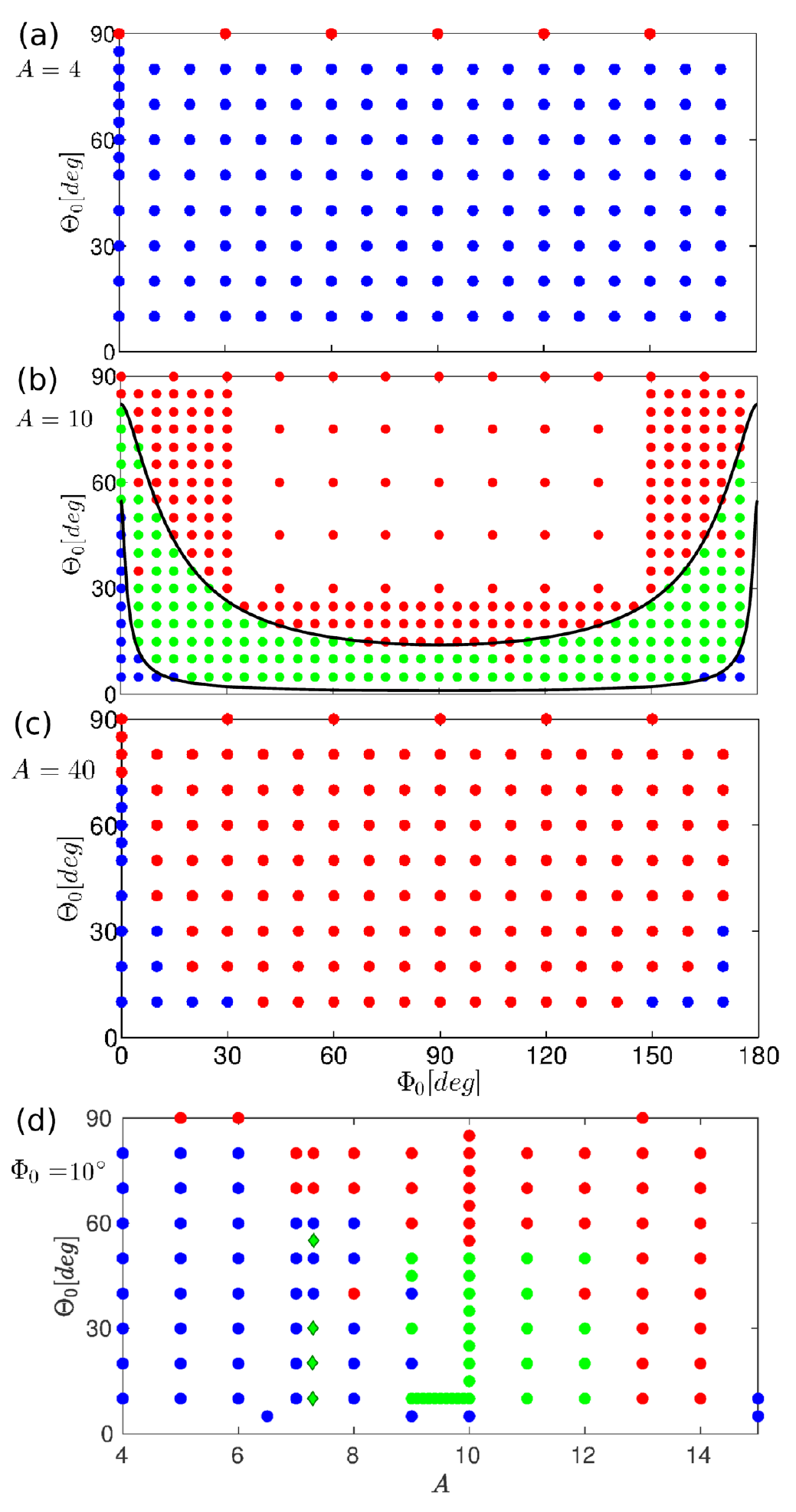}\hspace{0.7cm}

\vspace{-0.85cm}
\caption{Modes of flexible fiber dynamics for different initial orientations $(\Phi_0,\Theta_0)$ and different values of the  
bending stiffness ratio~$A$; (a) $A$=4, (b) $A$=10, (c) $A$=40, (d) $\Phi_0=10^{\circ}$. The color indicates convergence to a specific attractor: rolling (blue), tumbling (red), periodic meandering (green dots) or periodic squirming (green diamonds for $A$=7.3).} \vspace{-0.35cm}
\label{diag}
\end{figure}

In this paper, we have %also 
systematically investigated how the evolution pattern depends on 
the fiber's 
initial orientation $(\Phi_0,\Theta_0)$ for different values of the bending stiffness ratio $A$. A summary of the
most important findings is 
illustrated in Fig.~\ref{diag}(a-c) for $4 \le A \le 40$. Flexible fibers with $A\!=\!4$, shown in Fig.~\ref{diag}(a), belong to the rolling  mode (blue dots)
for all the initial orientations other than in the shear plane $xz$ ($\Theta_0\!=\!90^{\circ}$). For rather stiff fibers with $A\!=\!40$, shown in Fig.~\ref{diag}(c), 
the tumbling mode (red dots) dominates for most of the initial orientations, except some of those that are very close to the vorticity direction and lead to the rolling  mode (blue dots). 

For fibers of a moderate stiffness $A\!=\!10$, shown in Fig.~\ref{diag}(b),
there exists a range of the initial orientations that correspond to the periodic mode, with the same attracting periodic meandering orbit.  %marked as green dots. 
This range, marked  in Fig.~\ref{diag} by green dots, 
%For $A$=9, 11 and 12 the attractors are close to periodic. 
is approximately 
contained between two black solid lines. Their meaning will be explained in the next section by comparing with the Jeffery orbits \cite{Jeffery}. %....... ......... The initial orientations are chosen 
Close to the boundaries that separate different dynamical modes in Fig.~\ref{diag}, some irregularities appear and a relatively small change of initial orientations can result in a different dynamical mode. An example of such behavior for three close 
initial orientations was shown in Fig.~\ref{panel_t35}. This sensitivity is related to chaotic properties of the transition between the modes. 
 
We have analyzed the fiber dynamics for a wide range of values of the bending stiffness $A$. 
The periodic meandering solutions are observed for a certain range of $A$. In addition, for values of $A$ not far from this range, close-to-periodic meandering motions are observed. Examples of such motions are shown and discussed in Appendix \ref{B}. % for values of $A$ relatively close 

In Fig.~\ref{diag}(d), the dependence of the attracting solutions on the bending stiffness $A$ is shown for the initial orientations with $\Phi_0=10^{\circ}$ and different values of $\Theta_0$. %,  is shown in Fig.~\ref{diag}(d). 
The attracting periodic or close to periodic meandering motions (green dots) are visible for 
$9 \le A \le 12$. Examples of close to periodic meandering solutions are shown in Appendix \ref{B}. %the Supplemental Material in Figs.~12-13.)
We observe that for $A=7.3$, some of the initial orientations are attracted to another three-dimensional periodic orbit, corresponding to the squirming solution (marked by green diamonds). This finding illustrates that the squirming periodic motion can be transient or attractive, depending on the value of the bending stiffness ratio $A$.
%for two different values of the parameter $C$, i.e.  $C=0.245$ and $C=0.025$, and with the effective aspect ratio $r_e=29$ determined earlier for the tumbling mode \cite{Slowicka_chaos_2015}.

The results shown in Fig.~\ref{diag} indicate that there exist well-defined ranges of the fiber bending stiffness in which different solutions dominate. This result is significant for dilute suspensions of flexible fibers where hydrodynamic interactions between fibers are weak. In particular, our results (Fig.~\ref{diag}a) predict the existence of an ordered phase with all (or almost all) the fibers straightened out and parallel to each other (and to the vorticity direction), for a narrow range of values of the ratio A of bending to hydrodynamic forces ($4 \le A < 6$ for $N=40$). This range can be reached (or avoided) by adjusting the shear rate. Therefore, it should be possible to control the ordering of flexible fibers in dilute systems.

%\newpage $\;$ \newpage
%\newpage $\;$ \newpage
\vspace{-0.1cm}
\section{Comparison with Jeffery orbits}\label{coJ}
In this Section, we will demonstrate that the rolling and tumbling modes can be interpreted as effective Jeffery orbits of the flexible fiber end-to-end vector ${\bf n}$ (see Fig. 1(b) for the notation), with an effective amplitude that depends exponentially on time. We will also show that the meandering mode is essentially different from the Jeffery solution.
For the sake of clarity, 
we will focus on a fixed value of the bending stiffness $A=10$. The results can be easily generalized for a wider range of values. 
\subsection{Jeffery orbits}
To compare, we will first remind a reader of the classical Jeffery equations of motion of a rigid prolate spheroid with an aspect ratio $r$, oscillating periodically  in shear flow \cite{Jeffery,Graham}. We will keep the same parametrization of the spheroid orientation $(\Theta_J(t),\Phi(t))$ as for the end-to-end vector of the flexible fiber, and assume for simplicity that $\tan \Theta_J(0)\!=\!C_0,\;\Phi(0)\!=\!90^{\circ}$. The period of the motion, in units of $\dot{\gamma}$, is %given as
\beq 
T= 2\pi (r+1/r), \label{r}
\eeq %of the Jeffery's orbit 
and the evolution of the orientation angles satisfies %the following equations,
\begin{eqnarray}
%C = \tan \Theta(t) \sqrt{\sin^2 \Phi(t)+\frac{\cos^2 \Phi(t)}{r_e^2}},\label{JJ}
 \tan \Theta_J(t) \!&\!=\!&\! \frac{C_0}{\sqrt{\sin^2 \Phi(t)+\cos^2 \Phi(t)/r^2}},\label{JJ}\\
 \tan \Phi(t) \!&\!=\!&\! \frac{1}{r} \cot\left(\frac{rt}{1+r^2}\right). \label{cottan}
\end{eqnarray}
We will use the relation, which follows from Eq.~\eqref{cottan},
\beq
\cos^2 \Phi (t)=\frac{r^2 \sin\left(\dfrac{rt}{1+r^2}\right)}{r^2 + (1-r^2)\cos\left(\dfrac{rt}{1+r^2}\right)}.\label{cos2}
\eeq
%
%Motion of a prolate spheroid in shear flow The classical Jeffery's o
\subsection{Rolling and tumbling modes}
In this section, we will briefly recall basic features of Fig.~\ref{panel1}(a-b) and then analyze them 
both qualitatively and quantitatively, in a more general context. The characteristic variables and parameters in the rolling and tumbling modes will be labeled by (r) and (t), respectively.
For both %the rolling and tumbling 
modes, the time dependence of the fiber end-to-end orientation angle $\Theta(t)$, shown in 
Fig.~\ref{panel1}(a-b), illustrates two important features. First, the characteristic oscillation time  is almost constant in time. Second, the amplitude of the oscillations changes monotonically with time. 

\begin{figure}[h]
 \includegraphics[width=8.6cm]{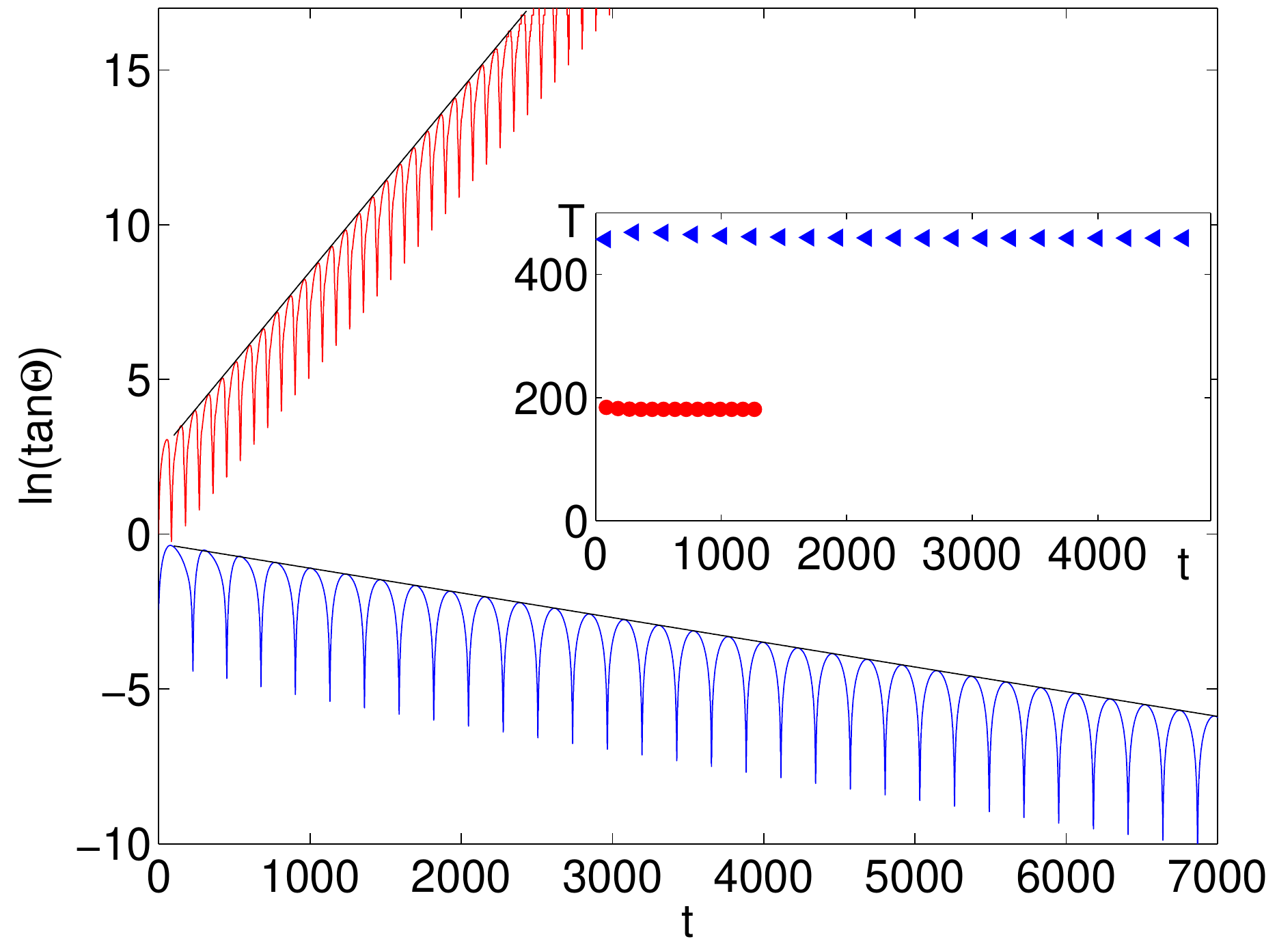}
 \caption{Oscillations of %the orientation angle
 $\tan \Theta(t)$ are approximately periodic %with period $T$ and 
 with an amplitude that changes exponentially with time. Main panel, lower curve (blue online): rolling mode with  $(\Theta_0,\Phi_0)$=$(5^{\circ},10^{\circ})$; upper curve (red online): tumbling mode with  $(\Theta_0,\Phi_0)$=$(45^{\circ},30^{\circ})$. The slopes of the solid straight lines are 
 $-t/1253 $ and $t/170 $, respectively. 
 Inset: Characteristic time $T$ of the oscillations is almost time independent. 
 $T(t)$ is the time difference between the $i$-th and $i+2$-th maxima, and the $i$-th maximum is at time $t$. $T_r$: triangles (blue online); $T_t$: dots (red online). %$T(t) is defined as Maxima of the function $\theta$ occur at Characteristic time of the oscillations
 }\label{Ttau}
\end{figure}

We highlight important features of the dynamics in Fig.~\ref{Ttau}.
In the inset we confirm that the oscillation time is practically time independent, with 

\begin{subeqnarray}
&&T\equiv T_r= 459 \hspace{0.95cm} \mbox{for 
%-- to 
the rolling mode,\hspace{0.5cm}}\label{Tr}\\% and $
&&T\equiv T_t= 181\hspace{1cm} \mbox{for 
%-- to 
the tumbling mode,\hspace{0.5cm}}\label{Tt}
\end{subeqnarray}
respectively.~In the main panel of Fig.~\ref{Ttau}, 
we plot $\ln |\tan \Theta|$ versus time and obtain %almost perfectly 
linear dependence of the maxima and minima on time. In this way 
we demonstrate that for the rolling mode, the amplitude of $\tan \Theta$  decays exponentially while for the tumbling mode, the amplitude of $\tan \Theta$  grows exponentially with time. The characteristic relaxation times are 
\begin{subeqnarray}
&&\tau \equiv \tau_r = - 1253 \hspace{0.5cm}  %correspond to 
\mbox{for the rolling mode}, \hspace{0.5cm}\label{taur}\\%and the sign `+' 
%and 
&&\tau \equiv \tau_t=170 \hspace{1cm} \mbox{for 
%-- to 
the tumbling mode,\hspace{0.5cm}} \label{taut}\end{subeqnarray}
%Both features are 
as illustrated in Fig.~\ref{Ttau} by straight lines of the corresponding inclinations. 
We have checked that the values of the characteristic oscillation times $T$ and the relaxation times $\tau$ are practically the same for other initial conditions leading to the same mode. 

Moreover, we will now demonstrate that 
\beq
\tan \Theta(t)= \exp( t/\tau)\, \tan \Theta_J(t), \label{idea} %\hspace{0.3cm}\mbox{where} \hspace{0.3cm}\theta_r(t)=\theta_r(t+T),
\eeq
with the periodic function $\Theta_J(t)$, which will be matched to the  Jeffery solution given by Eq.~\eqref{cottan}, and negative or positive values of $\tau$ given by Eq.~\eqref{taur}.%-\eqref{taut}. 
%%%%In the exponent, the sign `-' and 
%$\tau \equiv \tau_r = 1253$  correspond to the rolling mode, and the sign `+' and $\tau \equiv \tau_t=170$ 
%-- to the tumbling mode.
%the fitted paramters 
%\bee
%$%T_r= 459, 
%\tau \equiv \tau_r = 1253$ for the rolling mode and 

%and almost periodic function $\theta \equiv \theta_r(t)=\theta_r(t+T)$. %\label{rolling_par}
%\eee
%The values of the parameters are practically the same for all the initial conditions leading to the rolling mode. 

%Very similar behavior is observed for the tumbling mode. Again, the characteristic oscillation time $T\equivT_t = 181$ is almost constant in time and the amplitude of the oscillations decays expenentially with time, now with a much shorter relaxation time $\tau \equiv \tau_t=170$, as described by the relation $90^{\circ} \!- \!\Theta(t) = \exp(-t/\tau) \,\theta(t)$ with $\theta \equiv \theta_t(t)=\theta_t(t+T)$. These properties of the rolling and tumbling modes are illustrated in Fig.~\ref{Ttau} where behavior of $\Theta$ is analyzed at consecutive times $t_{max}$ when the function $\theta$ has a local maximum of value $\theta_{max}$.  

%Therefore, it is worthwhile to check if the rolling dynamics can be interpreted as a dumped effective Jeffery orbit. 
Matching the period $T$ of $\Theta_J(t)$, given by Eqs.~\eqref{Tt}, %-\eqref{Tr}, 
with the Jeffery period of a spheroid with the aspect ratio $r$, given by Eq.~\eqref{r}, %\cite{Jeffery},
%\beq T= 2\pi (r+1/r), \label{r} \eeq %of the Jeffery's orbit 
we obtain (for $A$=10) $r\!=\!r_r\!\!= \!73.0 $ as `the effective aspect ratio' of the flexible fiber in the rolling mode. Note that this value is significantly larger than the geometrical aspect ratio of the deformed fiber (the spring constant $k$ was chosen to be so large that the fiber is practically inextensible). Moreover, the shape of the fiber significantly changes in time, while the period of the damped oscillations and `the effective aspect ratio' remain almost constant in time,  decreasing only by a few percent.
%.... 2 \% but check other init cond .... %during the fiber evolution
%\subsection{Tumbling mode}
 For the tumbling mode, the effective aspect ratio following from Eq.~\eqref{r} is much shorter, $r\!\!=\!\!r_t\!\!=\!\! 28.8$\footnote{The tumbling motion in the shear plane of the end-to-end vector of a flexible fiber  was matched with an effective Jeffery orbit in Ref.~\cite{Slowicka_chaos_2015}.}.
 %This value of $r$ was determined by matching  the period $T$=181 of the tumbling motion for $A$=10 with the Jeffery's period  of ``an effective spheroid''\cite{Slowicka_chaos_2015}, \eqref{r}}.

\begin{figure}[b!]
\vspace{-0.3cm}
\hspace{-0.4cm}  \includegraphics[width=8.25cm]{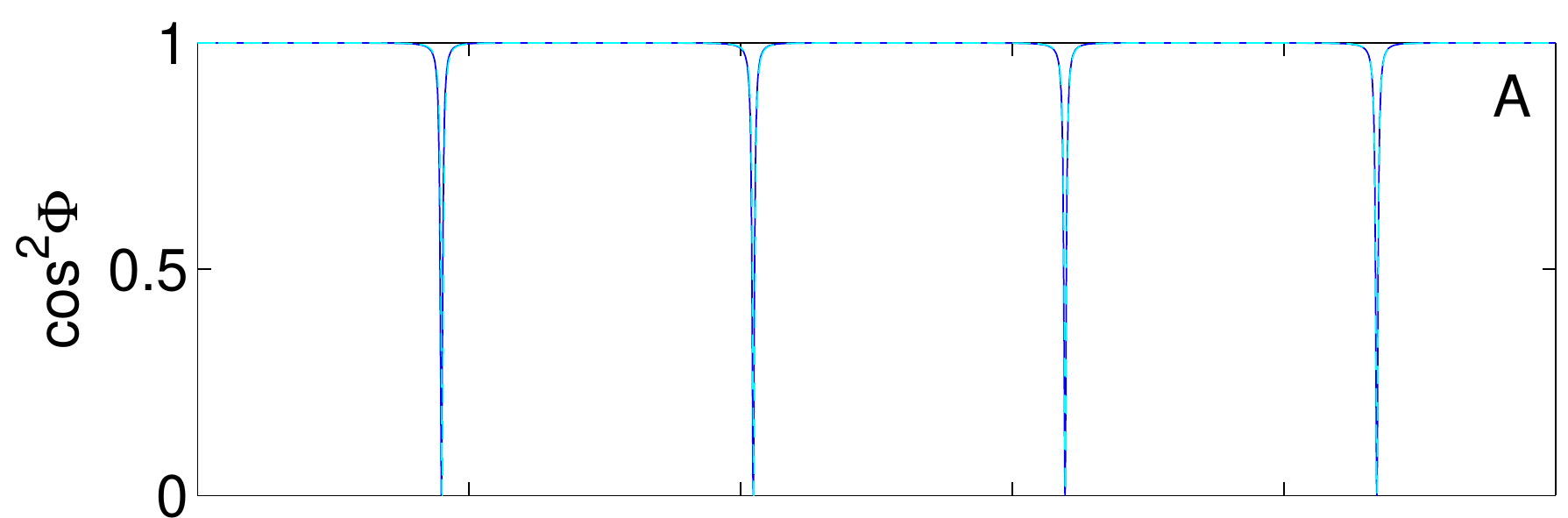}\\
 \includegraphics[width=8.6cm]{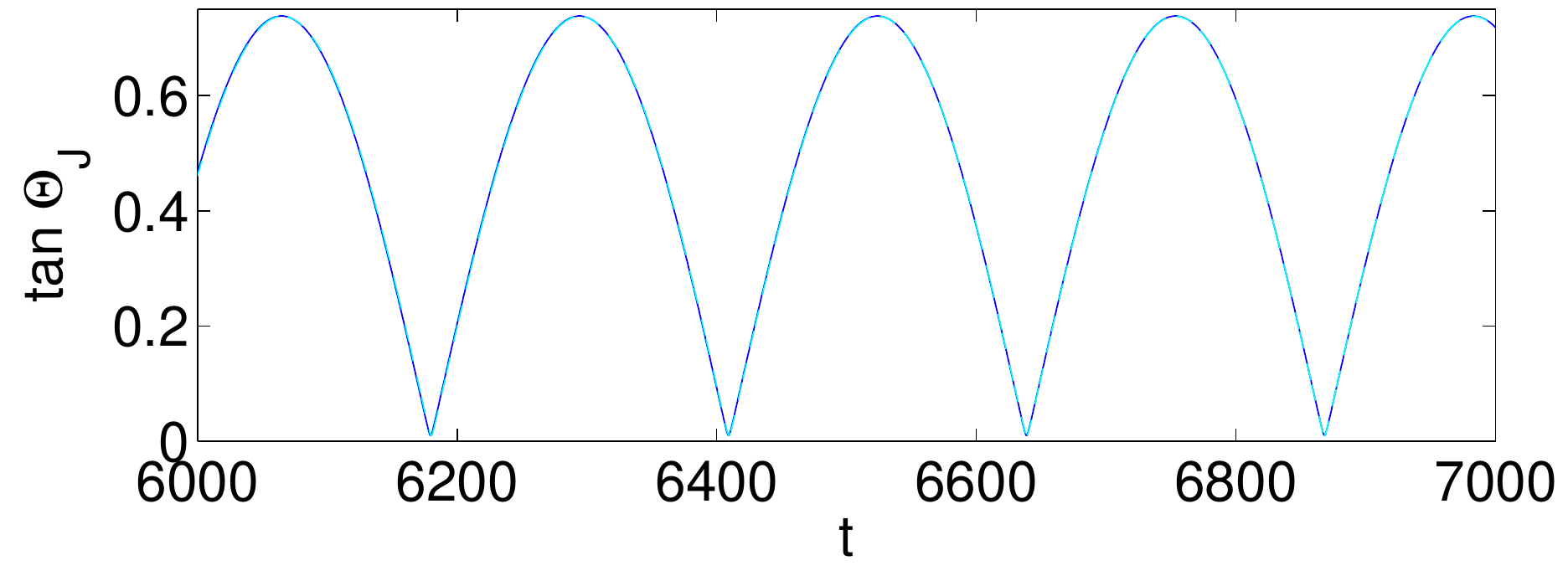}\\ %\vspace{-0.9cm]
\hspace{-0.5cm}  \includegraphics[width=8.3cm]{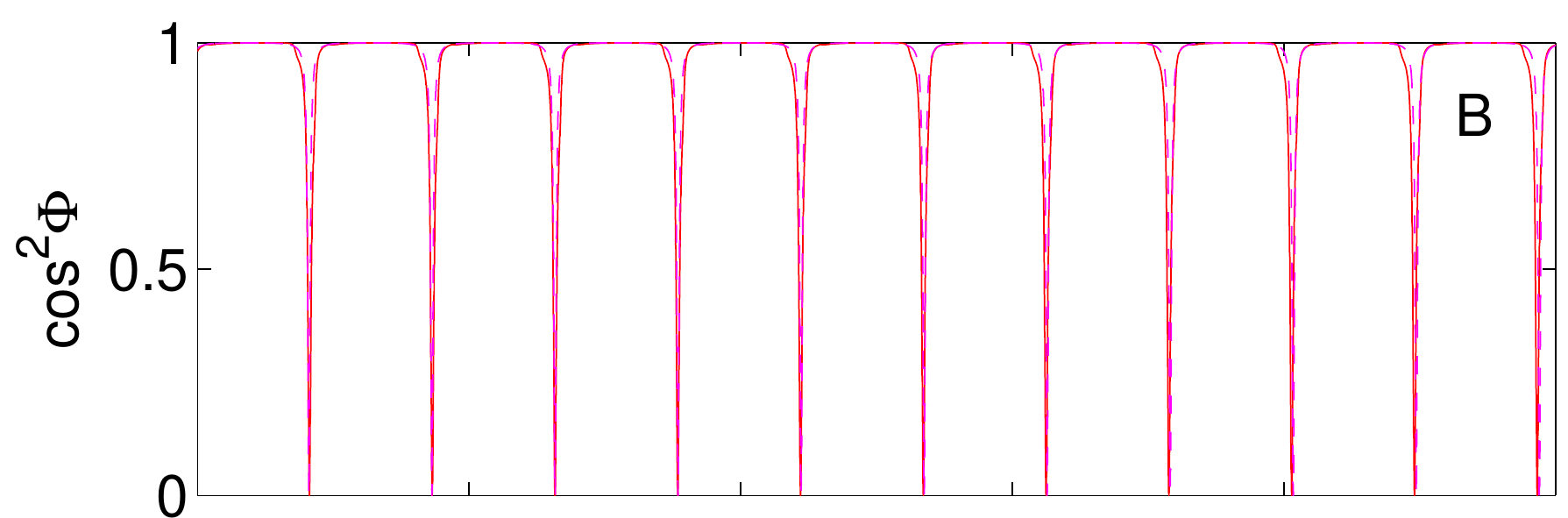}\\
 \includegraphics[width=8.6cm]{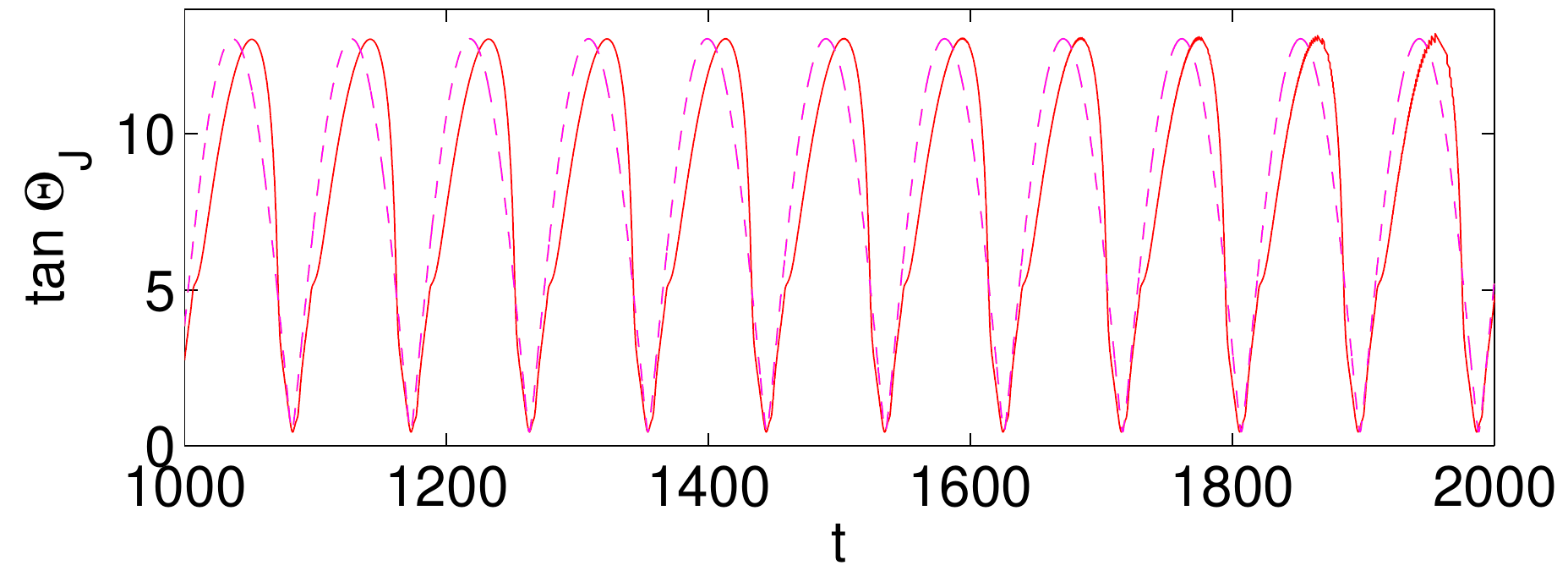}
 \vspace{-0.6cm}
  \caption{Rolling (A) and tumbling (B) modes interpreted by Eq.~\eqref{idea} as effective Jeffery solutions. A: $\Theta_0$=$5^{\circ}$, $\Phi_0$=$10^{\circ}$, $C_0$=$0.0101$.  %$\tan\Theta_J$=$\tan\Theta\exp(-t/\tau$), $t_J$=$t$-5\\
  B: $\Theta_0$=$45^{\circ}$, $\Phi_0$=$30^{\circ}$, $C_0$=$0.463$. % $\tan\Theta_J$=$\tan\Theta\exp(t/\tau$), $t_J$=$t$-13. 
  Solid lines: numerical results. Dashed lines: Jeffery approximation given by Eqs.~\eqref{JJ} and~\eqref{cos2} with $r\!=\!r_r=73.0$ and $r\!=\!r_t=28.8$.  
 %black: C=0.25 and C=0.0195. %$T(t) is defined as Maxima of the function $\theta$ occur at Characteristic time of the oscillations
 }\label{compare} \vspace{-0.2cm}
\end{figure}

 In Fig.~\ref{compare} we compare $\Theta_J(t)$=$\exp(- t/\tau)\tan \Theta(t)$ and $\cos^2\Phi(t)$ evaluated numerically (solid lines) with the %the oriention angles following from the 
 effective Jeffery expressions~\eqref{JJ} and~\eqref{cos2} (dashed lines), in which time is shifted back (by 5 and 13 units) to match the initial conditions and to compensate for transients, including the small decrease of $T$ with time in the first stage of the evolution. Also, for the same reason, %in the plots 
 we use the fitted values of $C_0$ given in the caption of Fig.~\ref{compare}. They slightly differ from the corresponding initial values:  %of $C(t)$ -- 
$C(0)$= 0.0152 for the rolling mode and $C(0)$=0.501
and for the tumbling mode. %$C(0)$=%$C_0$=

For the rolling mode, the numerical and analytical curves are superimposed. For the tumbling mode, there appear some differences, but still $\Theta_J(t)$ can by approximated remarkably well by the Jeffery solution. 
%$C_0*\sqrt{r^2-(r^2-1)\cos\left(t/\tau \right)}$$\tau=r/(1+r^2$
% tann=C0./sqrt(1/r^2+(1-1/r^2)*(cos(b1(:,3)*pi/180)).^2);

In this way, we arrive to one of the main conclusions of this paper. The 
rolling and tumbling modes can be interpreted as `effective Jeffery orbits'. The word `effective' means that the standard expressions \eqref{JJ}-\eqref{cos2} for the Jeffery trajectories still hold, but now with a constant value of the parameter $C_0$ replaced by  the time dependent amplitude $C(t)$,
\beq
C(t)= C_0 \exp(t/\tau),\label{Ct} %\hspace{0.3cm}\mbox{where} \hspace{0.3cm}\theta_r(t)=\theta_r(t+T),
\eeq
with the negative and positive values of $\tau$ corresponding to two different attractors of the dynamics.
The decay of $C(t)$ to zero for the rolling mode and the growth to infinity for the tumbling mode describe, respectively, the rates of convergence to the rolling stationary state and to the periodic two-dimensional tumbling restricted to the shear plane.%two different attracting solutions.

\begin{figure}[h]
 \includegraphics[width=8.6cm]{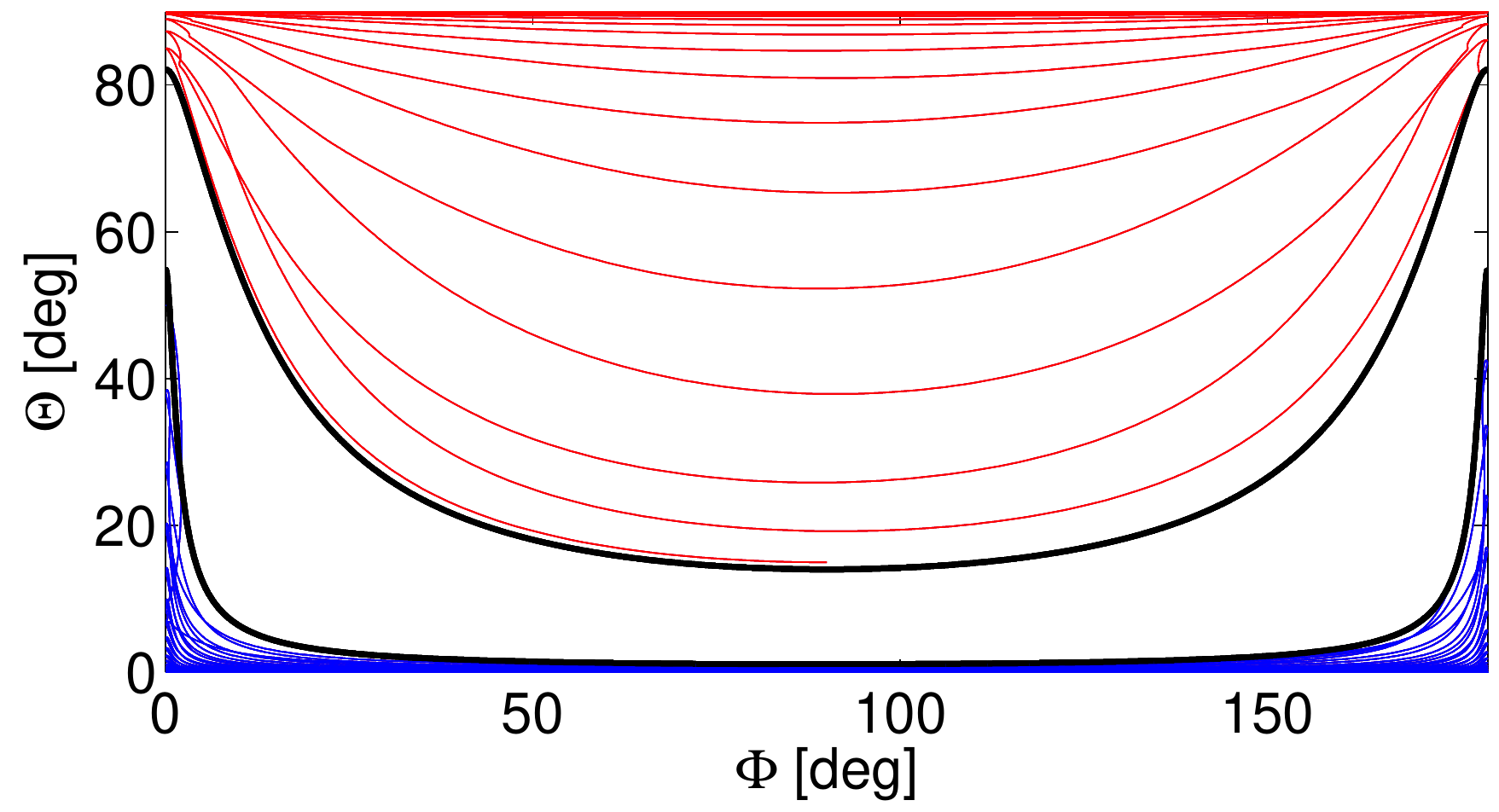} \vspace{-0.3cm}
 \caption{Rolling (blue) and tumbling (red) modes evolving towards $\Theta=0^{\circ}$ and $\Theta=90^{\circ}$, respectively from two different initial conditions.
 Red: $\Theta_0=15^{\circ}$, $\Phi_0=90^{\circ}$, $C_0$=0.268. Blue: $\Theta_0=50^{\circ}$, $\Phi_0=0^{\circ}$, $C_0$=0.0163. Black: Jeffery solutions with $C_0$=0.0195, $r%\!=\!r_r
 $=73.0 (bottom) and $C_0$=0.25, $r%\!=\!r_t
 $=28.8 (top). %are plotted also in Fig.~\ref{comparison_Jeffery} with thick black lines.
 %and $C_0$=0.0195. %$T(t) is defined as Maxima of the function $\theta$ occur at Characteristic time of the oscillations
 }\label{dziura}
\end{figure}

To analyze the two  different patterns of the evolution, and inspect the transition between them, we investigate other initial conditions. In particular, in Fig.~\ref{dziura} we choose %correspond to 
the initial values of the amplitude $C$ much closer to each other, i.e. 
$C_0$=0.0163 and $C_0$=0.268. 
The different time dependence of $C$ in both modes is clearly visible in Fig.~\ref{dziura} 
where we use the simulation results to plot $\Theta$ versus $\Phi$. In this plot, the change of $C$ can be traced along the trajectories as the consecutive values of $\arctan \Theta$ at $\Phi=90^{\circ}$. 
For different initial conditions with $C_0>$0.25 and $C_0<$0.0195, the evolution in time is qualitatively the same as shown in Fig.~\ref{dziura}. In the next subsection, we will discuss the dynamics for 0.0195$<C_0<$0.25, which corresponds to the empty space that separates the rolling and tumbling modes in Fig.~\ref{dziura}.

\subsection{Meandering modes}
\begin{figure} [h!] 
%\vspace{-0.7cm}
\vspace{-0.1cm}
\hspace{-1cm} \includegraphics[width=9.8cm]{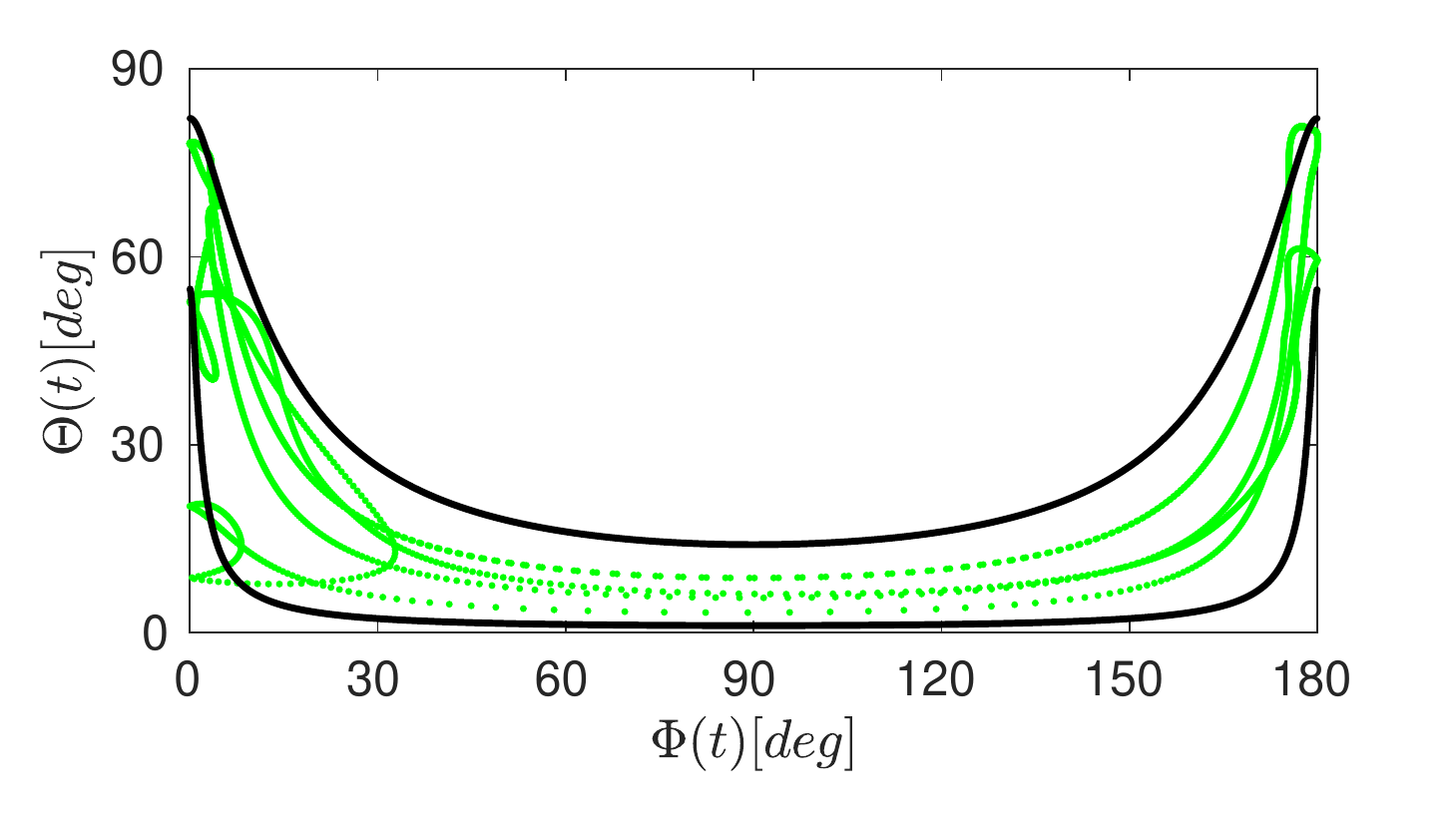}\hspace{0.3cm}\\
\vspace{-0.3cm}
\caption{Periodic meandering solution (green dots) compared to Jeffery orbits (black solid lines) of a rigid effective spheroid with $C\!=\!0.25$, $r\!=\!28.8$ (top) %$C\!=\!
and %$C\!=\!0.0163$, 
$C\!=\!0.0195$, 
$r\!=\!73.0$ (bottom). Here $A=\!10$, $\Theta_0$=10$^{\circ}$, $\Phi_0$=10$^{\circ}$ and 19164$\le\!t\!\le$19531.%; see equation \eqref{r}.\vspace{0.5cm}
} 
\label{comparison_Jeffery}
\end{figure}

The essential property of the periodic meandering solution is that it differs significantly from the Jeffery orbits \cite{Jeffery}. % of an effective spheroid. 
 In Fig. \ref{comparison_Jeffery}, we plot (in green) $\Theta(t)$ versus $\Phi(t)$ for the meandering mode, evaluated numerically for $\Theta_0=10^{\circ}$, $\Phi_0=10^{\circ}$. The corresponding initial value of $C$ is $C_0$=0.0307, with $r\!=\!r_m$=117, as evaluated from Eq.~\eqref{r} for $T_m$=735. It is clear that the meandering trajectory is much more complicated than the %an effective
 periodic Jeffery orbit. In this case there is no systematic drift of the trajectory, and therefore the concept of `an effective Jeffery orbit' cannot be applied.
 %
%Here we demonstrate this difference by comparing the evolution of the end-to-end vector with the motion of a rigid spheroid with an effective aspect ratio $r_e$. The value of $r_e \approx 28.8$ is determined by matching  the period $T$=181 of the tumbling motion in the tumbling mode  for $A$=10 with the Jeffery's period  of ``an effective spheroid''\cite{Slowicka_chaos_2015}, \eqref{r}
%\beq T= 2\pi (r_e+1/r_e). \eeq 
%In Fig. \ref{comparison_Jeffery}, we present (in green) the time-dependent relation between the angles $\Phi(t)$ and $\Theta(t)$ that determine the instantaneous orientation of the end-to-end vector ${\bf n}$ (see Fig. 1(b) for the notation). 
In Fig. \ref{comparison_Jeffery}, 
%For comparison 
we also plot %the same black lines as 
two black solid lines that follow from 
%In Fig. \ref{comparison_Jeffery}, it is shown that the periodic solution $g_1$ (green) differs significantly from Jeffery orbits, which are plotted for comparison (black solid lines), with the use of 
the Jeffery relation \eqref{JJ} as in Fig.~\ref{dziura}. The meandering trajectory is almost everywhere located inside the region determined by these lines. 
%\bee
%C = \tan \Theta(t) \sqrt{\sin^2 \Phi(t)+\frac{\cos^2 \Phi(t)}{r_e^2}},\label{JJ}
%C = \tan \Theta(t) \sqrt{\sin^2 \Phi(t)+\cos^2 \Phi(t)/r_e^2},\label{JJ}
%\eee
%for two different values $C=0.25$ and $C=0.0195$.
%\vspace{-0.5cm}
%

\indent We investigated numerically evolution of flexible fibers with different initial orientations. For most of the initial conditions with 0.0195$<C_0<$0.25, we have found the meandering mode, which fills the empty space that separates the rolling and tumbling modes in Fig.~\ref{dziura}. 

\subsection{Physical interpretation}
The analysis of the effective Jeffery orbits and the meandering modes, performed in the previous subsections, provides a simple physical explanation for  %the results of our systematic analysis of 
the dependence of the modes on the initial orientation,  %illustrated
%shown 
determined in Fig.~\ref{diag}(b) for a moderate bending stiffness ratio $A$=10. In this diagram, the phase space of the initial orientation angles separates into three distinct regions, each of them leading to a different dynamical mode, and finally to a different dynamical attractor. The borders between these regions are separated by the two special Jeffery solutions with $C_{0}$=0.0195  and $C_{0}$=0.25, plotted as the black lines in Figs.~\ref{diag},~\ref{dziura} and~\ref{comparison_Jeffery}. The effective Jeffery solutions, convergent to the tumbling and rolling modes  exist for a sufficiently large and a sufficiently small value of $C_0$, respectively. However, they do not exist for a range of intermediate values of $C_0$. In this range, periodic or close to periodic motions dominate. %Fibers he black curves which

This generic classification of the modes is perturbed by the existence of transient, close to periodic motions for some of the orientations and bending stiffness ratios $A$, as illustrated in Figs.~\ref{panel2add}-\ref{panel_t35}.

In future work we hope to provide analytical justification for these new observations and quantitive results.

\section{Discussion and conclusions}
One of the main results of this paper is that 
there exist periodic and close to periodic three-dimensional motions of a flexible fiber in shear flow. 
The meandering periodic solution is an attractor for a certain range of the relative bending stiffness $A$. The squirming motion is a transient in this range, and an attractor in a narrow range of smaller values close to $A$=7.3. 
%We have also found other three-dimensional, transient or attracting periodic motions (which will be presented elsewhere). 
%
Our results indicate that %by changing the ratio $A$ of bending to shear forces their essential effect of  is the
a change of the relative bending stiffness $A$ may trigger a transformation between different periodic (or close to periodic) solutions, and the corresponding change of the characteristic {\it sequence} of three-dimensional fiber shapes.  
We have demonstrated that the complexity of a flexible-fiber shape (and in particular the number of local maxima of the curvature) may change significantly with time. 
A more detailed study of %different three-dimensional, transient or attracting 
periodic and close to periodic motions of flexible fibers in shear flow, for a wide range of the bending stiffness $A$,  will be presented elsewhere.
%in a periodic or close to periodic way.  %in a regular or almost regular way
%This is a new perspective ........
%In experiments, it should be easy would be to find 
%The squirming close to periodic solution %which 
%is typical at both short and moderate time scales, and therefore it may be %possible to be 
%observed %detected in experiments 
%more easily than the periodic meandering motion that dominates at long times. Actually, 

We have shown also that the time-dependent rolling and tumbling modes of a flexible fiber in shear flow
can be interpreted as effective Jeffery solutions, with the constant $C_0$ replaced by
an exponential function of time $C(t)$ given by Eq.~\eqref{Ct}. Therefore, the effective Jeffery orbits drift towards one of two attractors: the fiber aligned with the vorticity direction and the fiber performing periodic motions entirely in the shear plane.
There exist two thresholds for the
initial value of $C(t)$: small $C(0)$ lead to the rolling mode, intermediate $C(0)$ 
to the meandering mode, and large $C(0)$ to the tumbling one. These thresholds are sensitive to the bending stiffness ratio $A$. 

%In this study, we used 
Unlike the slender body theory \cite{duRoure,Liu}, the {\sc Hydromultipole} method used in this study takes into account the fiber thickness. This feature is important to study the three-dimensional dynamics in shear flow, and in particular, the effective Jeffery orbits and the tumbling motion, in contrast to the elastica models \cite{BeckerShelley,YoungShelley} that predict an infinite tumbling time of infinitely thin fibers. 

Finally, we comment on possible comparisons with experiments. 
In Refs.~\cite{Harasim2013,Liu}, quasi-2D trajectories of actin were investigated, with the focal plane of the motion perpendicular to the vorticity direction. 
%
%Although the fibers we study do not perform Brownian motion and are much shorter than actin, %investigated experimentally~\cite{Harasim2013,Liu}, but t
The elasto-viscous
number $\bar{\mu}=8 (L/d)^4/A \approx 2.2 \times 10^6$,  i.e., the largest value of $\bar{\mu}$ analyzed 
%in the experiments 
in Ref.~\cite{Liu}, 
corresponds to $A$=10 for our system. 
Indeed, for $A$=8 we recover a shape evolution similar to Movies 5 and 6 from \cite{Liu}, with U turns and S turns that appear irregularly. 
%, shown in Movie 6. %Although 
%However, %for so flexible fibers 
%This type of motion is exceptional rather than typical in our simulations of fibers with a similar bending stiffness. 

\begin{figure}[t] \vspace{-0.1cm}
 \includegraphics[width=8.6cm]{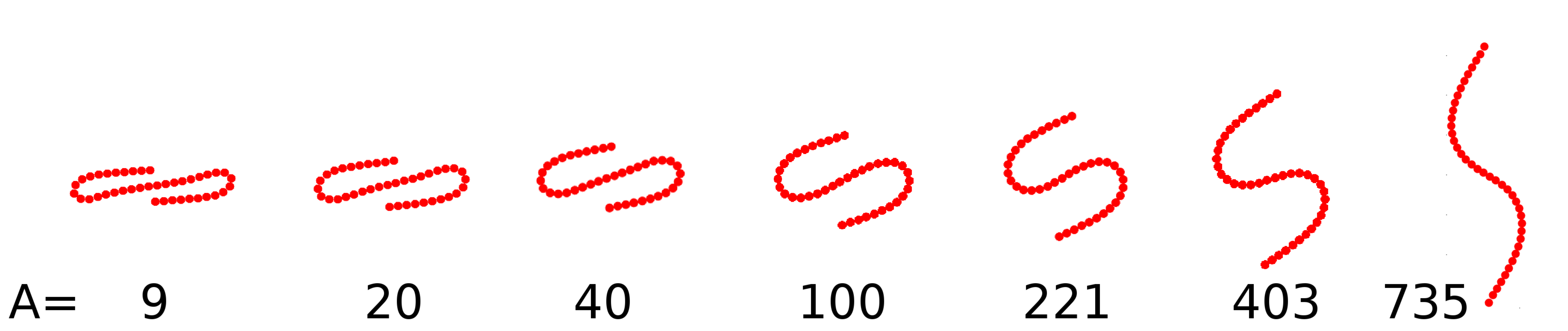} \vspace{-0.3cm}
 \caption{Shapes of fibers which perform periodic tumbling motion are shown 
 at the flipping time, defined as the time when the end-to-end vector is perpendicular to the flow. The values of the bending stiffness $A$ are indicated. Initially, $(\Theta_0,\Phi_0)=(90^{\circ},0^{\circ})$. A detailed study of the quantitative features of the fiber shapes as a function of $A$ will be published in the future. 
 }\label{S-shapes}
\end{figure}

However, we observe that more stiff fibers, after a long time, approach 
the periodic tumbling motion, with the S-turns only,   as illustrated in Fig.~\ref{S-shapes}.
%see also Fig. \ref{curvature-stiff} in the Supplemental Material. 
These shapes are different from the C-turns observed in \cite{Liu}. A 
more detailed study is needed to understand the reason for such a difference. The obvious differences are that the fibers 
we study do not perform Brownian motion and are much shorter than actin. Moreover, in Ref. \cite{Liu}, the simulations were performed in 2D and the experiments were carried out in a bounded geometry. Finally, we note 
that even a very small curvature of the external flow can have a drastic influence on flexible fiber shapes \cite{farutin}.
%, but this comparison indicates that it should be possible to observe periodic motions in experiments. 
%indicates that such ty
%our findings resemble the experimental observations of actin from Refs.~\cite{Harasim2013,Liu}. ............describe differences ......... 
%However, this comparison is incomplete because - as far as we know - there is no measurements of the $xy$ projection. 
%In both cases, the fiber is typically very deformed and thin in the z direction.
%Therefore, t
%

As a last remark, we suggest searching for three-dimensional, periodic motions in experiments. Our estimates indicate that they appear in the range of the bending stiffness accessible in experiments. 
With this goal in mind, we highlight the need to perform video recordings in the $xy$ (flow-vorticity) plane where the characteristic repeatable evolution of shapes should be visible. We hope to report the results of such experiments in a future communication. \vspace{-0.1cm}

%\section  
%\newpage
\acknowledgments
%\section{ACKNOWLEDGMENTS}
\noindent 
 %M.L.E.J. %acknowledges scientific benefits from COST Action MP1305 and 
We thank Olivia du Roure for helpful discussions. A.M.S. and M.L.E.J. were supported in part by %the National Science Centre 
Narodowe Centrum Nauki under grant No.~2014/15/B/ST8/04359.  H.A.S. thanks the NSF CMMI-1661672. We benefited from the ITHACA project PPI/APM/2018/1/00045 financed by the Polish National Agency for Academic Exchange.

%{\color{green} Additional plots to be done: three projections of the time-dependent trajectories of the unit vector: 
%$(\delta x/\Delta L, \delta y/\Delta L, \delta z/\Delta L)$ and at the same plot, two time-dependent trajectories of the unit vector aligned with the rigid spheroid which performs the Jeffery orbit at
%$ C=0.245$ and $C=0.025$, respectively.}
%It is clear that these orbits significantly differ from each other. 

\appendix
%\newpage
\section{Periodic meandering attractors}\label{A}
In this section, we provide more details about the periodic meandering solution. %In Fig.~\ref{panel1}, we have illustrated that, in 
%%%%%%%In both cases, the 
%this Letter we focus on 
%The fiber almost straightens while tumbling. 
In Fig.~\ref{panel1}(c) the maximal values of the relative length of the end-to-end vector are close to one. In the figure, by looking at the orientation $\Theta(t)$  we observe that during the periodic meandering motion, the fiber straightens while tumbling 
at an orientation that is not along the flow. This property is also visible in Fig. \ref{shapes} where fiber shapes are shown. 
In particular, for the periodic meandering solution with $A$=10, the fiber is almost straight for $\Phi$=0 and $\Theta\!\approx \!78^{\circ}$. %This is different than in case of 
In contrast, for  the tumbling solution, the fiber straighten along the flow, i.e. at $\Theta = 90^{\circ}$. 
%the fiber is almost straight, %with $\Delta L/L_0 \approx 1$
%see inset of Fig.~\ref{panel1}(c). Inset: 3800$<t<$4800. ................

\begin{figure} [t!] 
\vspace{-0.2cm}
%\hspace{0.8cm}$A=100, \Theta_0=5^o, \Phi_0=10^o$\hspace{1.6cm} $A=100, \Theta_0=60^o, \Phi_0=60^o$ \hspace{1.6cm} $A=10, \Theta_0=10^o, \Phi_0=10^o$\\
\includegraphics[width=6.5cm]{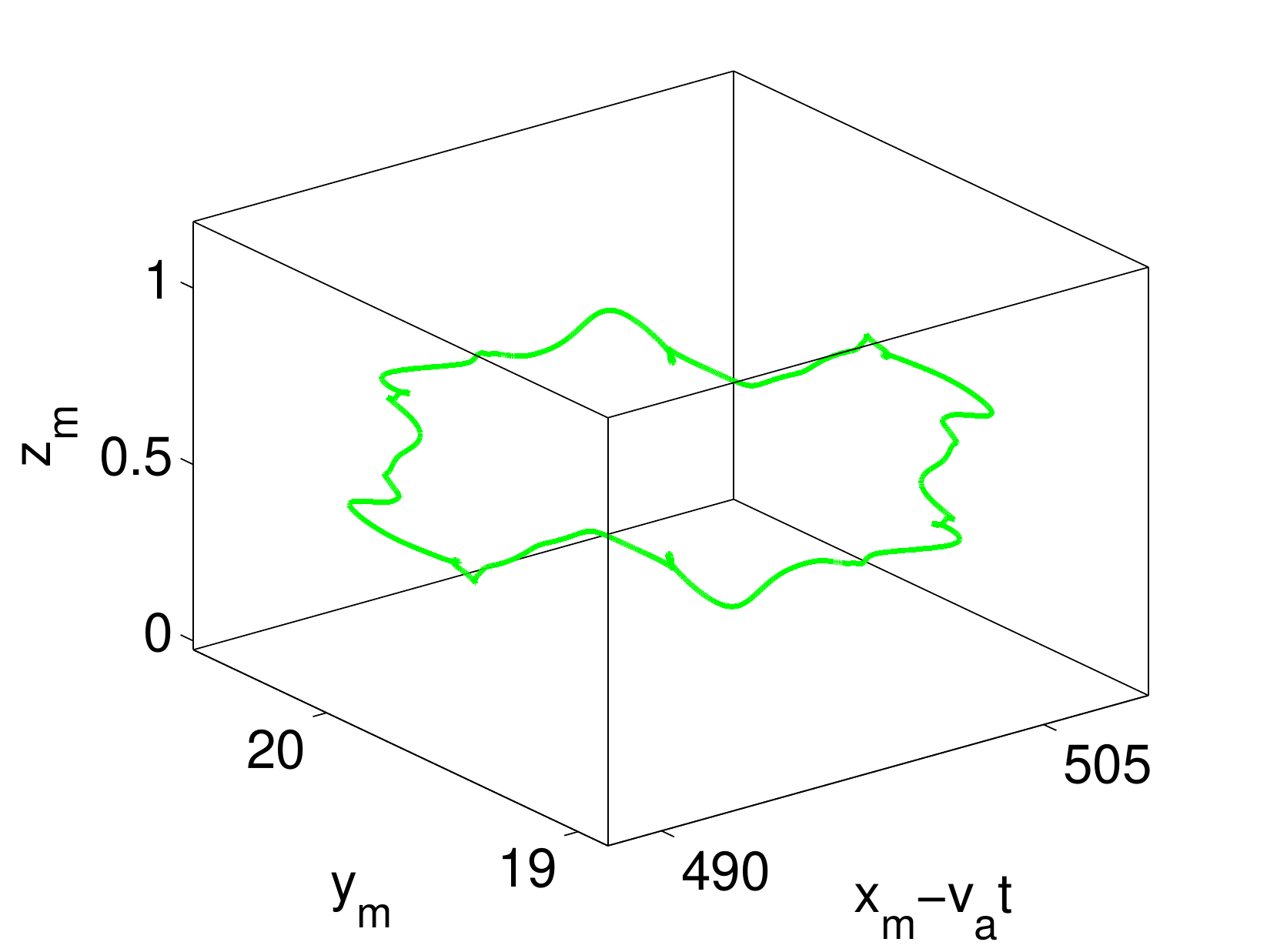}
\vspace{-0.32cm}
\caption{Fiber center-of-mass periodic trajectory for $A\!=\!10$. 
%$ 18796 \le t \le 19531$
%Typical trajectories of the end-to-end vector for three modes of the fiber dynamics: blue, red and green. {\color{green} Remove titles of frames in the second and third rows (the titles in the first row are sufficient)}
%Top: the time dependent angle $\Theta$. Bottom: the instantaneous relative length $\Delta L/L_0$ of the fiber end-to-end vector. %up- $\Theta$ evolution in time; middle - $\Delta L/L_{0}$ - the relative length of the end-to-end vector of the fiber in time; bottom - the Jeffery orbit parameter $C$ -  evolution in time. Left column - blue mode: results for $A=100, \Theta=5^o, \Phi=10^o$. Central column - red mode: results for $A=100, \Theta=60^o, \Phi=60^o$. Right column - green mode: results for $A=10, \Theta=10^o, \Phi=10^o$.
%{\color{green} Change A=100 into A=10.}
} \vspace{-0.3cm}
\label{center-of-mass}
\end{figure}

In the meandering motion, every bead performs a periodic orbit superposed with translation along the shear flow with a velocity {\bf v}$_{\text a}$, equal to the mean velocity of the fiber center-of-mass, averaged over the period.
%In the frame of reference translating along the flow with the mean velocity $v_a$ of the fiber center-of-mass (averaged over the period), every bead perfoms a periodic orbit. 
%\subsection{Periodic attracting solution $g_1$ for $A$=10}
%For $A\!=\!10$, the period $T\!=\!735
%367
The periodic meandering motion of the center-of-mass for $A\!\!=\!\!10$ is shown in Fig.~\ref{center-of-mass} and Movie~4, in the frame of reference translating along the shear flow with  velocity {\bf v}$_{\text a}$. Here, $A=10$, and the initial conditions are the same as in Figs.~\ref{panel1}, \ref{panel2add}, \ref{nn}, \ref{shapes} and Movie~2. The time range~is $99188\!\! \le \!\!t \!\!\le \!\!99988$, with the period  $T_m\!\!=\!\!735$. 
Owing to symmetry, trajectories of the end-to-end vector, displayed in Fig.~\ref{nn}, %panel2add}-\ref{panel_t35}, 
close after $T/2$. Our results show that the meandering period is longer than the transient squirming period and 
much longer than the tumbling period. % (which for $A$=10 are equal to $T\approx 305$ and $T=181$, respectively). 

%We will now discuss properties of the periodic solution, and show that it significantly differs from the Jeffery orbits. 
%
%The periodic solution is superimposed with translation along the flow with a constant speed ${v}_{cm}$. We evaluate the center-of-mass position $({x}_{cm},{y}_{cm},{z}_{cm})$ of the fiber and determine the average center-of-mass translation speed %$\mbox{v}_{\mbox{a}}$ along the ambient flow,
%\bee
%{v}_{cm}= \frac{{x}_{cm}(t+T)-{x}_{cm}(t)}{T}. \label{average_velocity}
%\eee
%We have verified that in the frame of reference translating along $x$ with speed ${v}_{cm}$, every bead performs periodic motion. For $A\!=\!10$, the period $T\!=\!735
%367 $.
%

\section{Close-to-periodic meandering attractors}\label{B}%: meandering solutions}
\begin{figure} [b!] 
\vspace{-0.3cm}
%\begin{center}
%{\bf \normalsize Quasiperiodic motion}
%
%\vspace{0.11cm}
%\hspace{0.8cm}$A=100, \Theta_0=5^o, \Phi_0=10^o$\hspace{1.6cm} $A=100, \Theta_0=60^o, \Phi_0=60^o$ \hspace{1.6cm} $A=10, \Theta_0=10^o, \Phi_0=10^o$\\
%5.7
%\psfrag{$\delta z$}{$\delta z/L_0$}
\includegraphics[width=5.62cm]{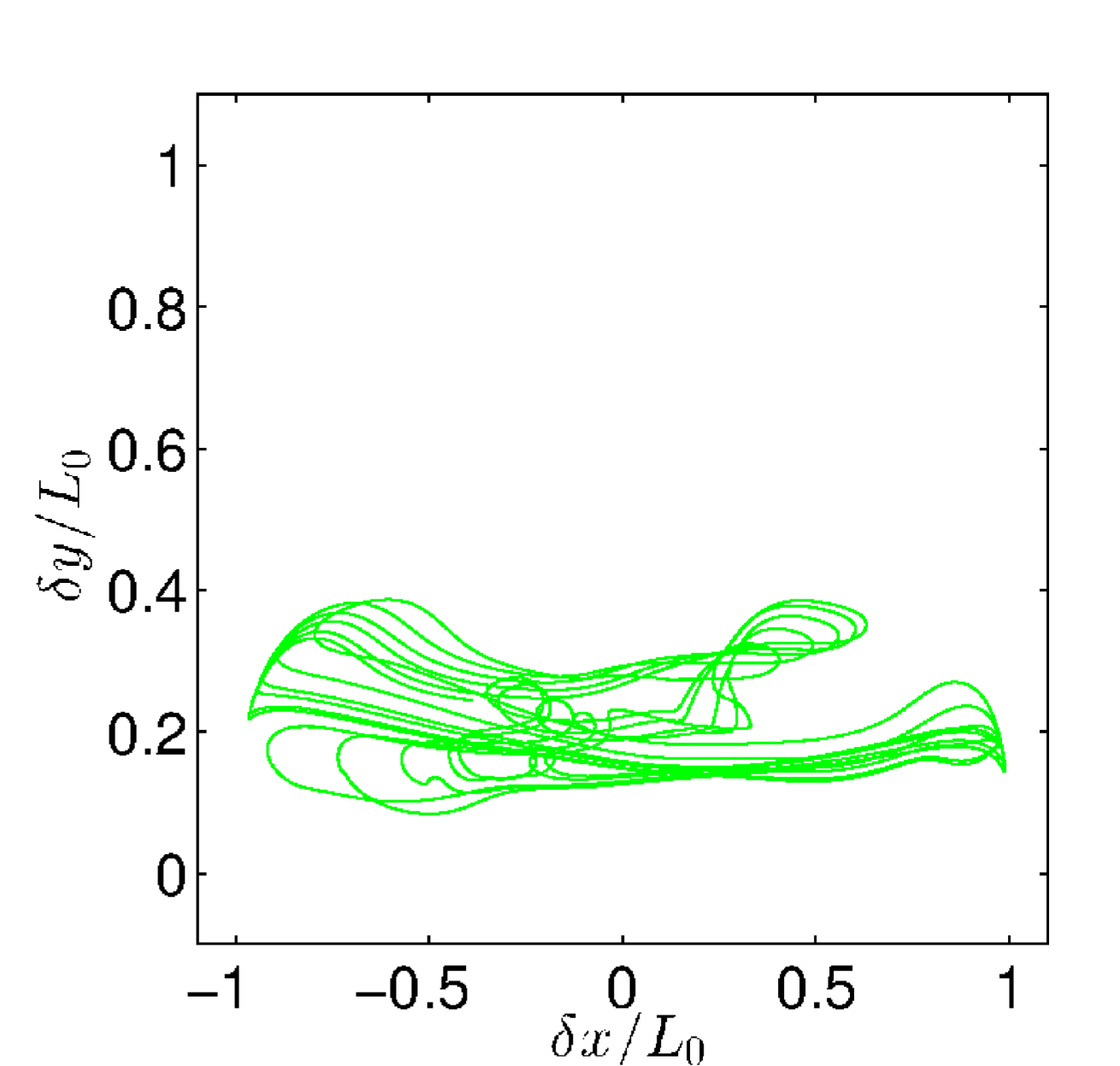} \vspace{0.05cm}
\includegraphics[width=5.62cm]{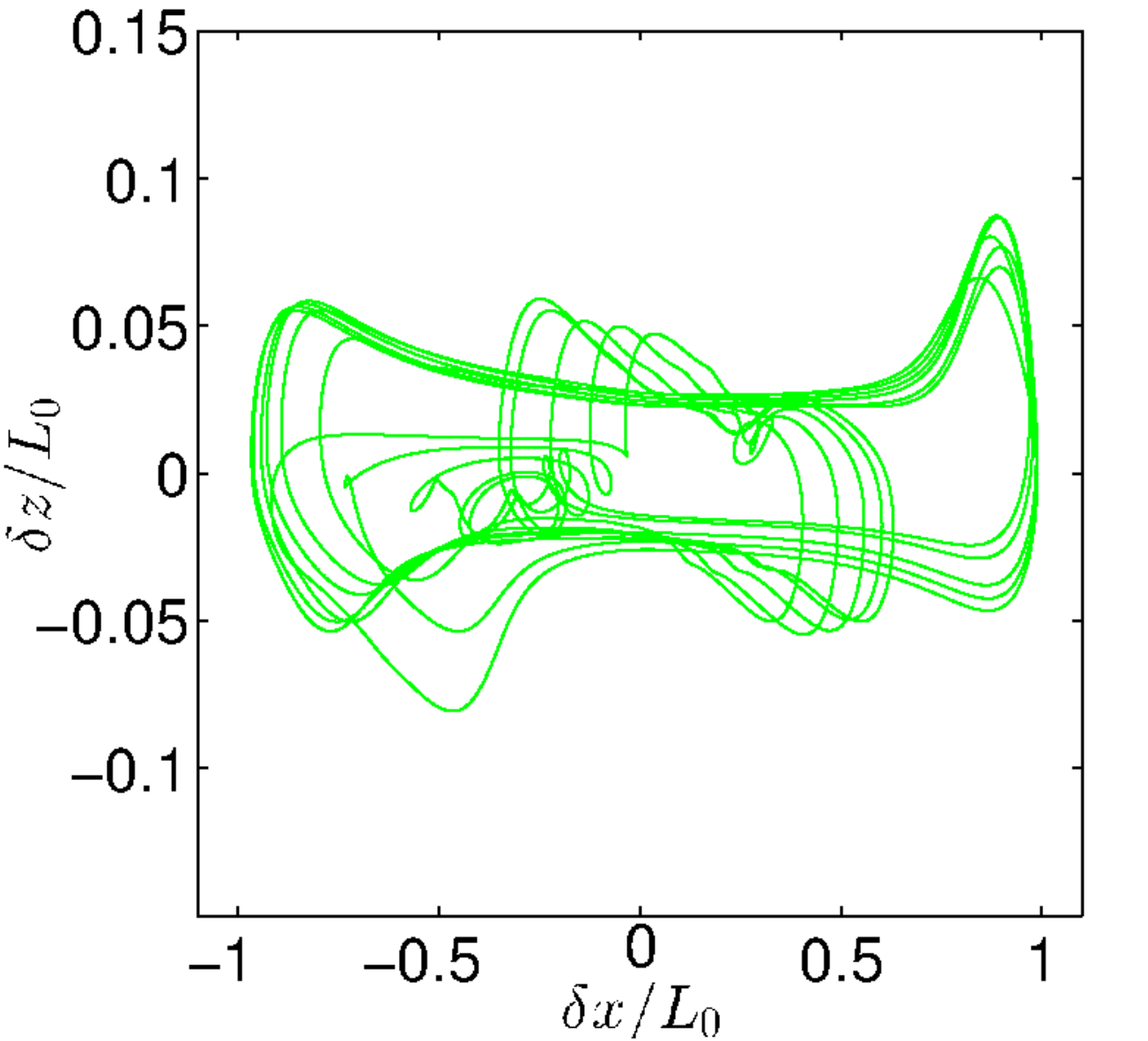} \vspace{0.08cm}
\includegraphics[width=5.62cm]{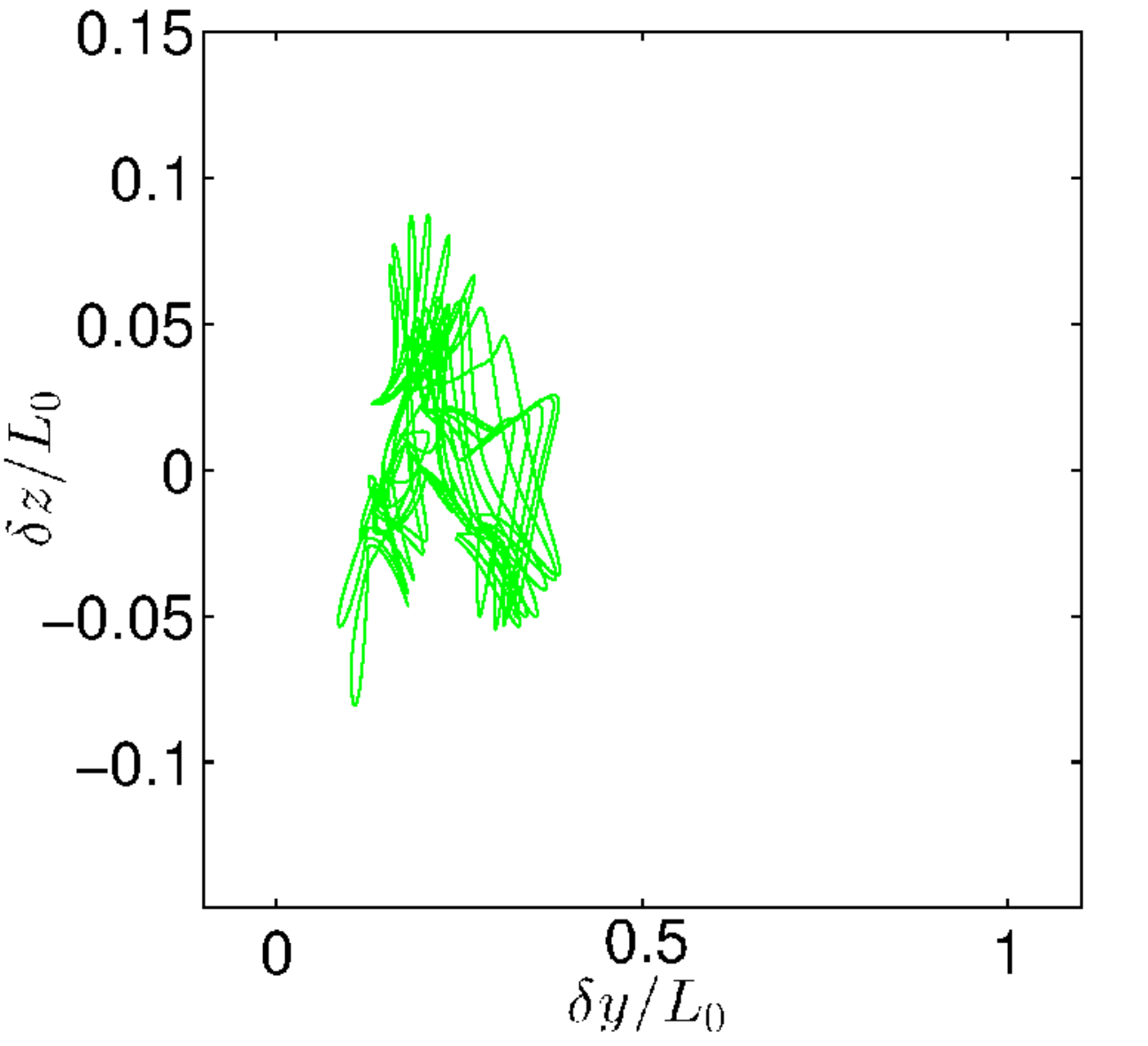}
\vspace{-0.4cm}
\caption{Example of close to periodic meandering motion: trajectories of the fiber end-to-end vector. Here $A$=11, $\Theta_0$=55$^{\circ}$,  $\Phi_0$=0$^{\circ}$, and the time range is $10565\le t \le 12565$. \\\\%The initial orientation as indicated. 
%Typical trajectories of the end-to-end vector for three modes of the fiber dynamics: blue, red and green. {\color{green} Remove titles of frames in the second and third rows (the titles in the first row are sufficient)}
%Top: the time dependent angle $\Theta$. Bottom: the instantaneous relative length $\Delta L/L_0$ of the fiber end-to-end vector. %up- $\Theta$ evolution in time; middle - $\Delta L/L_{0}$ - the relative length of the end-to-end vector of the fiber in time; bottom - the Jeffery orbit parameter $C$ -  evolution in time. Left column - blue mode: results for $A=100, \Theta=5^o, \Phi=10^o$. Central column - red mode: results for $A=100, \Theta=60^o, \Phi=60^o$. Right column - green mode: results for $A=10, \Theta=10^o, \Phi=10^o$.
%{\color{green} Change A=100 into A=10.}
} 
\label{qp}
%\end{center}
\end{figure}

In Fig. \ref{qp} we illustrate that by changing values of the bending stiffness $A$ away from $A$=10, we observe close to periodic (rather than periodic) meandering motions, which are long-lasting and vary similarly to the periodic meandering solution for $A$=10.
For $A$=9, departure from periodicity has a different pattern: %it is clear that 
the system oscillates back and forth between the twin meandering orbits, symmetric with respect to $\delta x \rightarrow -\delta x$ and $\delta z \rightarrow -\delta z$.

The end-to-end periodic meandering trajectory, shown in Fig. \ref{nn}, is non-symmetric. Naturally, after rotation by %

\newpage
\onecolumngrid
\bc
\begin{figure} [b!] \vspace{-0.65cm}
\includegraphics[width=18.2cm]{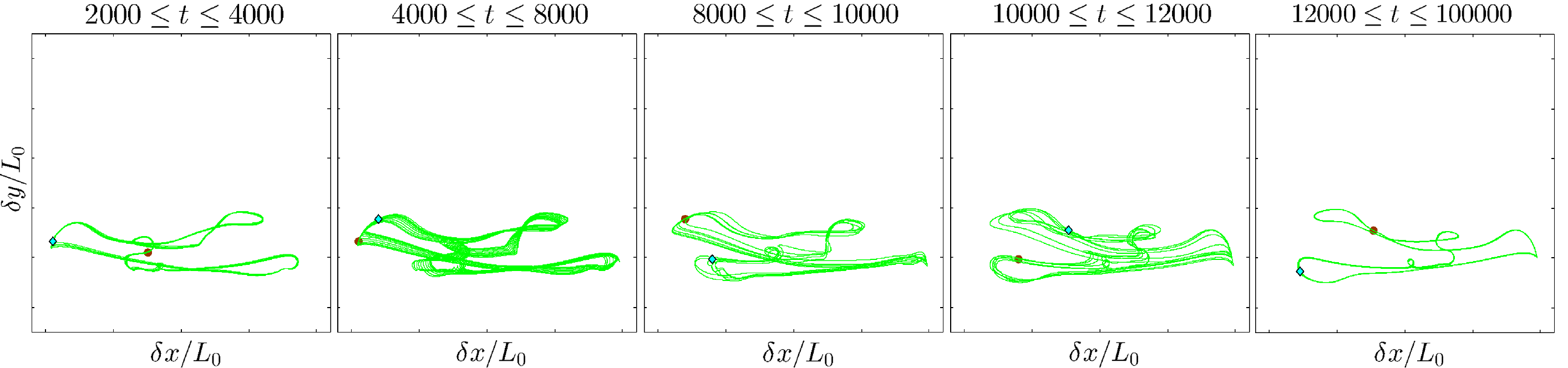} \vspace{-0.5cm}
\caption{Long-time effects for periodic meandering mode with $A$=10: trajectories of the fiber end-to-end vector. %Brown 
Dots (brown online) and %cyan 
diamonds (cyan online) mark 
positions, respectively,  at the beginning and at the end of the indicated time range.\vspace{0.3cm}}\label{sym1sym2}
\end{figure}
\ec
%\twocolumngrid
%
%
%Moreover, we observed long time effects of flipping from one to the other trajectory, as shown by the 
\begin{figure}[h!]
%\hspace{-0.76cm}
\vspace{-0.2cm}
\hspace{-0.55cm}
\includegraphics[width=15.6cm]{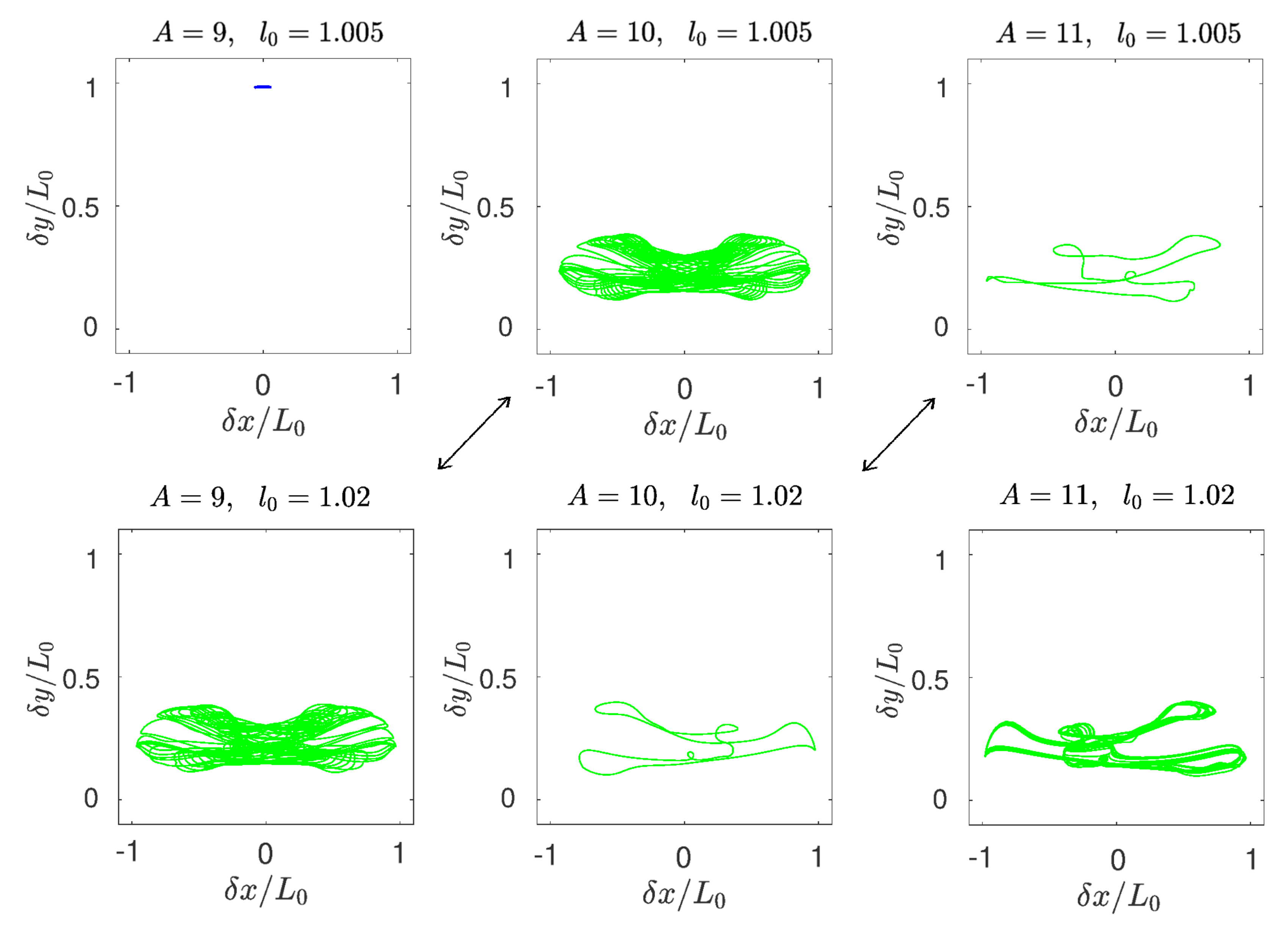}
%Fi13.eps} 
\vspace{-0.7cm}
%figures/fig_panel_2.eps}
\caption{The long-time trajectories of the fiber end-to-end vector. % reached after very large times. 
Comparison between fibers with different values of  $l_0$ and $A$ (as indicated),  %. Examples are shown 
for the same initial orientation  $(\Theta_0,\Phi_0)=(10^{\circ},10^{\circ})$. The plots are made 
for  $ 5000 \le t \le 10000$ (upper row) and 
%Comparision between different models of $l_0$ distances.  of the initial end-to-end vector ${\bf n}_0$ and coresponding flexibilities $A=9,10,11$ for two values $l_0$. Trajectories are plotted for the last $5000$ time units of the simulations. 
%(lower row) Trajectories are plotted 
for 
$ t_{end}-5000 \le t \le 
t_{end}$ (lower row), where $t_{end}=5\cdot 10^4$ for A=9 and 11, and $t_{end}=10^5$ for $A=10$. See also Table 
%new: (lower row) A=9 and 11: $t_{end} = 50000$, and for A=10, $t_{end}=10^5$.
\ref{tab1}.\\}
\label{fig1sm} 
\end{figure}

\twocolumngrid
\noindent 
$\pi$ around the $y$ axis, one obtains another periodic orbit.   
Both appear in our simulations (compare Figs.~\ref{panel2add}-\ref{panel_t35} and \ref{nn}). 
Moreover, we observed long time effects of flipping from one to the other trajectory, as shown by the example in Fig.~\ref{sym1sym2} (again, for $\Theta_0=\Phi_0=10^{\circ}$).

%(compare the first and the last frames in Fig. \ref{sym1sym2}). 
%
%In Fig.~\ref{J}, we plot the evolution of the end-to%, and with the effective aspect ratio $r_e=29$ determined by matching  the period of the tumbling motion in the red mode $T$=181 for $A$=10 with the Jeffery's period $T= 2\pi (r_e+1/r_e) $ %of the Jeffery's orbit of ``an effective spheroid''\cite{Slowicka_chaos_2015}.
%
%determined earlier for the tumbling mode \cite{Slowicka_chaos_2015}.
%
%...... panel with Jeffery orbits of the ened-to-end vector ...
%

%.......Finally, it is worth to display examples of certain long-time effects. 
The periodic meandering mode for $A$=10 seems to be stabilized already for times around 2000. However, as shown in Fig.~\ref{sym1sym2}, there appear some long-time effects: the trajectory later flips to another one, symmetric with respect to rotations along the vorticity direction (i.e., with respect to $\delta x \rightarrow -\delta x$ and $\delta z \rightarrow -\delta z$), and later remains in this shape. 
%\newpage $\;$ \newpage
\section{Accuracy of the simulations}\label{C}
The Stokes flow dynamics of an elastic fiber modeled as a chain of beads, evaluted using the {\sc Hydromultipole} numerical algorithm, is more accurate for a larger multipole truncation order $LL$ 
and a smaller distance $l_0$ between the centers of consecutive beads. 
Numerical accuracy of the simulations of flexible fibers in shear flow, performed with the {\sc Hydromultipole} numerical codes \cite{cichocki}, was analyzed in detail in Ref. \cite{Slowicka_chaos_2015} for the periodic tumbling motion of fibers located in the $xz$ plane. %As shown there, 
The accuracy of the truncation at $LL\!=\!2$   was estimated there as 3\%, and %the accuracy 
for the $l_0\!=\!1.02$ approximation as 10-15\%. It is essential to keep the gap between consecutive bead surfaces small enough to prevent the beads from rotating under shear independently from each other. For small gaps, the lubrication effects serve this goal quite effectively. 

Similar accuracy estimates are expected to hold for the three-dimensional motions considered in this article. However, in this case, the analysis of the accuracy is more subtle because we have already observed that the existence of periodic meandering solutions is limited to a narrow range of the bending stiffness. % $A$ around $A=10$. %(and by analogy, a limited range of the distances between the bead centers $l_0$ is slso expected). 
A small departure from this range gives rise to motions that are close to periodic (for example, see Fig. \ref{qp}). 
Therefore, taking into account the chaotic nature of flexible fibers dynamics in shear flow \cite{Slowicka_chaos_2015}, 
the basic question is if three-dimensional periodic meandering motions are also observed for more precise (but computationally more demanding) bead models of a flexible fiber, with a smaller~$l_0$.
%equilibrium values of the distance $l_0$ between the consecutive bead centers. 

Therefore, we choose $A\!=\!9,\,11$ as the values of the bending stiffness close to $A\!=\!10$, which for  $l_0\!=\!1.02$ corresponds to the meandering attractor  with the period $T\!=\!735$. In Fig.~\ref{fig1sm} and Table \ref{tab1}
we compare the long-time solutions obtained for %several close values of 
%$A=9,\,10,\,11$ and 
$l_0\!=\!1.005$ and $l_0\!=\!1.02$. While changing $l_0$, we fix values of $N$ and $k$, and consider the identical initial orientations $\Theta_0\!=\!\Phi_0\!=\!10^{\circ}$.

As shown in Fig.~\ref{fig1sm} and Table \ref{tab1}, for $l_0\!=\!1.005$, the periodic meandering solution is not recovered 
at $A\!=\!10$; instead, for these values of the parameters, we find the close-to-periodic meandering motion, with a smaller maximal time $\tau\!=\!636$ 
between the consecutive tumblings. The decrease of the tumbling time for smaller values of $l_0$ was also observed for the tumbling periodic orbit \cite{Slowicka_chaos_2015}, and explained as the influence of the bead rotation. To minimize this effect, the gaps between the bead centers should be kept reasonably small. In practice,  $l_0\!\lesssim \!1.02$ is sufficient.

The long-time periodic meandering solution, detected for  $l_0\!=\!1.02$ and $A\!=\!10$ with the period $T\!=\!735$, 
is observed also for the  smaller distance $l_0\!=\!1.005$, but at a larger value of the bending stiffness $A\!=\!11$, and with a smaller 
period $T\!=\!653$. 
Movie 5 illustrates that the evolution of shapes in these two periodic solutions is almost identical (here time is normalized by the two different periods, respectively). We also 
%periods: $T\!=\!735$ for $l_0\!=\!1.02$ and $T\!=\!653$ for $l_0\!=\!1.005$.
%%%%%%similarity of the dynamics for $(A,l_0)=(9,1.02)$ to the dynamics for $(A,l_0)=(10,1.005)$,
%%%%%%%%%%and also similarity of the dynamics for $(A,l_0)=(10,1.02)$ to the dynamics for $(A,l_0)=(11,1.005)$.  
find that the long-time meandering quasi-periodic solutions 
for $l_0=1.02$ are similar to those obtained  for $l_0=1.005$, but with larger values of the bending stiffness~$A$, as shown in Fig.~\ref{fig1sm} and Table \ref{tab1}.
%
%as shown in Fig. \ref{fig1sm} and Table \ref{tab1}, where 
%
%Here we point out that the data from %the results from Ref. \cite{Slowicka_chaos_2015} indicate that the fiber dynamics evaluated for given values of $l_0$ and $A$ may correspond to the dynamics calculated for a smaller value of $l_0$ and a larger value of $A$, if $l_0$ is small enough.
%
%Following this idea, in this Letter 
%
%
%The initial position of a fiber was $\Theta_{0}=10^{o}, \Phi_{0}=10^{o}$ and parameters of elasticity $A=9,10,11$ for which the fiber tends to the attractor or to the almost periodic orbit in simulations 
%presented in the paper.
%Fig.~\ref{fig1sm} presents the long-time trajectories of the fiber end-to-end vector for different values of $l_0$ and $A$, and Table \ref{tab1} 
\begin{table}[h] \vspace{-0.4cm}
%\begin{center}
\caption{Long-time dynamics of fibers for different %values of 
$l_{0}$ \& $A$. }
\label{tab1}
\begin{tabular}{|p{0.78cm}|p{2.4cm}|p{2.4cm}|p{2.4cm}|}
\hline
$l_{0}\backslash A$ & 9 & 10 & 11\\
\hline
1.005 & %\mbox{\color{blue}  
rolling  &%{\text {\color{green}  
close to periodic
%}} {\text {\color{green} 
meandering
& %{\text {\color{green}  
periodic \;\;%}} {\text {\color{green}
meandering\\
\hline
1.02 & %{\text {\color{green} \small 
close to periodic 
%}} {\text {\color{green} 
meandering & %{\text {\color{green} 
periodic
%}} {\text {\color{green} 
meandering &  close to periodic meandering\\
\hline
\end{tabular}
%\end{center}
\end{table}
%describes the solutions reached after a long time. 
%Comparing the results, 

\vspace{-0.5cm}
\section{Description of the movies} %\vspace{-0.2cm}
\noindent {\bf Movie 1}. \\
%{\footnotesize \begin{verbatim} http://bluebox.ippt.pan.pl/~mekiel/fibers/Movie_1.avi\end{verbatim}}
Typical evolution of fiber shapes in the relaxation phase of the rolling (blue), tumbling (red) and meandering (green) modes. The bending stiffness $A=10$ and the initial orientations are the same as in the simulations shown in Fig. \ref{panel1}. The movie starts at the same time instants as the snapshots in Fig. \ref{panel1} but it lasts longer; 500 time units are shown for each mode (time is synchronized). Close to periodic squirming motion in the relaxation phase of the meandering mode is clearly visible.\\

\noindent {\bf Movie 2}. \\
%{\footnotesize \begin{verbatim} http://bluebox.ippt.pan.pl/~mekiel/fibers/Movie_2.avi\end{verbatim}}
%Typical shape evolusion of fibers attracted by: (left) the periodic 
%meandering mode and (right) the periodic tumbling mode. Movie corresponds
%with Fig. 2.
Typical  evolution of fiber shapes in the attracting periodic orbits:  meandering (green) and tumbling (red) solutions are reached after a long time. The bending stiffness $A=10$ and the initial orientations are the same as in Figs. \ref{panel1}-\ref{panel2add} and Movie 1. Time $t$ is synchronized, with 
$t_{0} \le t \le t_{0}+736$ where $t_{0}=99188$ for the meandering and  $t_{0}=4653$ for the tumbling solutions.
%Initially the fiber is straight with the orientations $(\Theta_0,\;\Phi_0)=(10^{\circ},10^{\circ})$ (left) and $(\Theta_{0},\Phi_{0})=(45^{circ},30^{circ})$ (right). The motion shown corresponds to long times as indicated in caption of Fig. \ref{shapes}(c).
%Periodic and close to periodic phases are reached after certain relaxation times. 
%Top: $xy$ projection; bottom: $xz$ projection. Snapshots from this move are shown in Fig. \ref{shapes}(a) and (b).
\\

\noindent {\bf Movie 3}. \\
%{\footnotesize{ \begin{verbatim} http://bluebox.ippt.pan.pl/~mekiel/fibers/Movie_3.avi\end{verbatim}}}
Illustration of the concept of the fiber end-to-end trajectories. 
A trajectory is drawn by the tip of the fiber end-to-end vector for the  periodic meandering solution. Here $A=10$, and the initial conditions are the same as in Fig. 3 and Movie 2. The time range is $99188 \le t \le 99988$.\\

\noindent {\bf Movie 4}. \\
%{\footnotesize{ \begin{verbatim} http://bluebox.ippt.pan.pl/~mekiel/fibers/Movie_4.avi\end{verbatim}}}
Periodic meandering motion of a flexible fiber center-of-mass, in the frame of reference translating along the shear flow with  velocity {\bf v}$_{\text a}$, equal to the mean velocity of the fiber center-of-mass, averaged over the period. Here, $A$=10. The corresponding trajectory is shown in Fig.~\ref{center-of-mass}.
\\ 

\noindent {\bf Movie 5}. \\
%{\footnotesize{ \begin{verbatim} http://bluebox.ippt.pan.pl/~mekiel/fibers/Movie_5.mpg\end{verbatim}}}
Testing the accuracy of our bead model. The sequence of fiber shapes in the periodic meandering long-time solution for $l_0\!=\!1.02$ and $A\!=\!10$ is almost the same as for a smaller (and more accurate) value $l_0\!=\!1.005$ and $A\!=\!11$. The time is normalized by the corresponding periods: $T\!=\!735$ for $l_0\!=\!1.02$ and $T\!=\!653$ for $l_0\!=\!1.005$.
%Movie 2.\\
%3D visualization of Movie 1.
%Meandering periodic (left), squirming close to periodic (middle) and tumbling periodic (right) motions of a flexible fiber in shear flow ${\bf v}=\dot{\gamma} z \hat{\bf e}_x$, for the bending stiffness $A=10$. Time is synchronized. 
%\vspace{4cm}

\end{document}